\documentclass[aps, prb , reprint]{revtex4-2}
\usepackage{verbatim}
\usepackage{bbold}
\usepackage{amsmath}
\usepackage[english]{babel}
\usepackage{graphicx}
\usepackage{mathtools}
\usepackage{hyperref} 
\usepackage{pdfpages}
\usepackage[normalem]{ulem}
\hypersetup{colorlinks=true, linkcolor=blue, citecolor=blue, urlcolor=blue}
\newcommand{\ket}[1]{|#1\rangle}
\newcommand{\bra}[1]{\langle#1|}

\makeatletter
\AtBeginDocument{\let\LS@rot\@undefined}
\makeatother

\begin{document}
\title{The role of polaron dressing in superradiant emission dynamics}

\author{J.~Wiercinski}
\email[]{jehw2000@hw.ac.uk}
\author{M.~Cygorek}
\author{E.~M.~Gauger}
\affiliation{SUPA, Institute of Photonics and Quantum Sciences, Heriot-Watt University, Edinburgh EH14 4AS, United Kingdom}
\date{\today}

\begin{abstract}
Cooperative effects of multiple quantum emitters are characterized by transitions via delocalized collective states with altered emission properties due to the existence of inter-emitter coherences. When realized with excitonic condensed matter nanostructures, these effects are significantly affected by the presence of strong emitter-phonon coupling, which leads to the formation of polarons. We show that, while for single-emitter emission into free space this formation has no impact on its radiative lifetime, the same is not true for superradiant emission. 
Considering the case of two indistinguishable quantum emitters, we analyse how polaron dressing affects collective photon emission by mixing bright and dark Dicke states.
Our numerical simulations show that this mixing crucially depends on the circumstances of the excitation of the system: Depending on the pulse length of an exciting laser, one can choose to either prepare polaronic Dicke states, or bare electronic Dicke states, changing the superradiant decay characteristics of the system. Additionally, we derive analytic expressions for these limiting cases, which match the results of numerically exact calculations.
\end{abstract}
\maketitle

\section{Introduction}
Non-classical inter-emitter correlations can fundamentally alter the emission of an ensemble of quantum emitters compared to a single one.
In an ideal system of optical dipoles, these correlations manifest in transitions along the `Dicke ladder' of symmetric multi-emitter states, while orthogonal parts of the Hilbert space are not accessed \cite{gross_superradiance_1982}. 
Paradigmatic for this is `Dicke superradiance' \cite{dicke_coherence_1954, gross_superradiance_1982}, where indistinguishability of dipoles results in enhanced decay rates leading to an overall superextensive scaling of optical emission. The enhanced emission from an ensemble of initially uncorrelated relaxed emitters has also been termed `superfluorescence' \cite{PhysRevA.11.1507}. Varying terminologies are discussed in Ref.~\cite{Cong:16}. 

Another related but distinct collective phenomenon is measurement-induced cooperative emission~\cite{skornia_nonclassical_2001, wolf_light_2020}, where emitters cannot be distinguished by the detectors. Then, due to the erasure of which-path information \cite{barrett_efficient_2005}, the wave function collapse after detecting a photon induces inter-emitter coherences that shape the emission profile and lead to two photon coincidence signals mimicking superradiance, even though the radiative decay rate remains unaffected \cite{koong_coherence_2022, cygorek_signatures_2023, wiercinski_phonon_2023}.

Cooperative quantum effects have high utility in quantum technologies as they can, for instance, mediate entanglement between distant qubits~\cite{barrett_efficient_2005}, be used to achieve collective strong coupling of organic molecules to cavities \cite{coles_strong_2014, coles_nanophotonic_2017}, support superabsorption \cite{higgins_superabsorption_2014,yang_realization_2021,quach_superabsorption_2022}, or serve as a heat reservoir in quantum engines \cite{kim_photonic_2022, kamimura_quantum-enhanced_2022}.

Solid-state platforms, such as quantum dots (QDs), have been established as sources of nonclassical light such as single photons \cite{senellart_high-performance_2017, PhysRevLett.123.017403,karli_super_2022}, entangled photon pairs \cite{huber_semiconductor_2018, PhysRevLett.123.137401, schimpf_quantum_2021, basset_quantum_2021}, and cluster states \cite{lindner_proposal_2009,istrati_sequential_2020}. Thus, they are promising candidates for realising collective quantum effects, as well.  
A crucial prerequisite to achieving superradiant emission, however, is a sample of indistinguishable quantum emitters, including in their spectrum and spatial positions, and even elaborate quantum dot growth techniques based on deterministic self-assembly \cite{laferriere_systematic_2021} typically produce quantum dots with significant variations in emission frequencies. Recently, the use of nanophotonic cavities or waveguides \cite{auffeves_few_2011, tiranov_collective_2023}, and sophisticated methods for tuning the resonance frequencies \cite{ding_tuning_2010, grim_scalable_2019} has made it possible to overcome these obstacles for InAs \cite{kim_photonic_2022}, and GaAs \cite{koong_coherence_2022} quantum dots.
Moreover, in perovskite materials superradiance has been observed, e.g. in nanocrystal superlattices \cite{blach_superradiance_2022}, while strong dipole enhancement in single nanocrystals has been attributed to single-photon superradiance \cite{zhu_single-photon_2024}. 

Solid-state emitters necessarily interact with their surrounding vibrational environment. In the single emitter case, this interaction leads to emission into phonon sidebands \cite{iles-smith_phonon_2017}. For cooperative emissions, on the other hand, it also leads to transitions away from the Dicke ladder. This effect, while generally reducing superradiance, can be utilized for quantum ratcheting \cite{werren_light_2023} and dark state protection \cite{PhysRevLett.111.253601,PhysRevLett.117.203603, rouse_optimal_2019}. While in the case of weak environment coupling the phonon influence can be described as a second-order perturbation to the system states leading to Markovian decoherence, in the strong coupling regime electronic and vibrational excitations mix and form vibronic excitations known as polarons \cite{hohenadler_lang-firsov_2007, nazir_modelling_2016}.

From a theoretical perspective, superradiant emission in the presence of a phonon environment poses a complex problem due to the involvement of multiple environments. Recently, non-additive effects between structured phonon and photon environments have been investigated \cite{hughes2011,rouse_optimal_2019,Denning_phonon_2020, rouse_analytic_2022,gribben_exact_2022}, and have become numerically tractable using numerically exact process tensor methods \cite{cygorek_simulation_2022, gribben_exact_2022, cygorek_sublinear_2023}.

In this article, we investigate superradiant emission from a pair of emitters focusing on the experimentally relevant situation of a strongly coupled vibrational environment, while the photon environment couples weakly and is unstructured. We find that the corresponding initial value problem can be solved analytically. Still, a non-trivial interplay between the environments emerges that can be understood by considering the dynamics of electronic and polaronic Dicke states, i.e.~Dicke states in the bare electronic basis and the polaron frame, respectively. We show that both types of Dicke states differ in their decay characteristics and can be optically addressed, depending on the excitation conditions. 

This article is organized as follows: First, in Sec.~\ref{sec:TLS} we consider a single two-level system (TLS) emitter interacting strongly with a vibrational bath and coupled to a photon environment with a flat spectral density.
We obtain, numerically as well as analytically, that the phonon environment leaves the radiative decay rate unchanged.
Afterwards, in Sec.~\ref{sec:2TLS}, we perform similar steps for a two-emitter system, obtaining renormalized rates for the polaronic dark and bright Dicke states. 
The dynamics of these two emitters, under different conditions, are the subject of the following Sec.~\ref{sec:results}: First, in ~\ref{sec:two_TLS_decoherence} we consider the phonon influence in the absence of radiative decay. 
Then, we consider excitation with a weak (Sec.~\ref{sec:2TLS_weak_excitation}) laser pulse and afterwards (Sec.~\ref{sec:2TLS_excitation_dependence}) investigate the dependence on the laser pulse area. We derive analytical expressions for the excitation number for both ultrashort and very long laser pulses and show how the pulse duration is connected to the excitation of polaronic and electronic Dicke states.
Finally, our results are summarized in Sec.~\ref{sec:summary}.

\begin{figure*}
    \centering
    \includegraphics[width=0.9\linewidth]{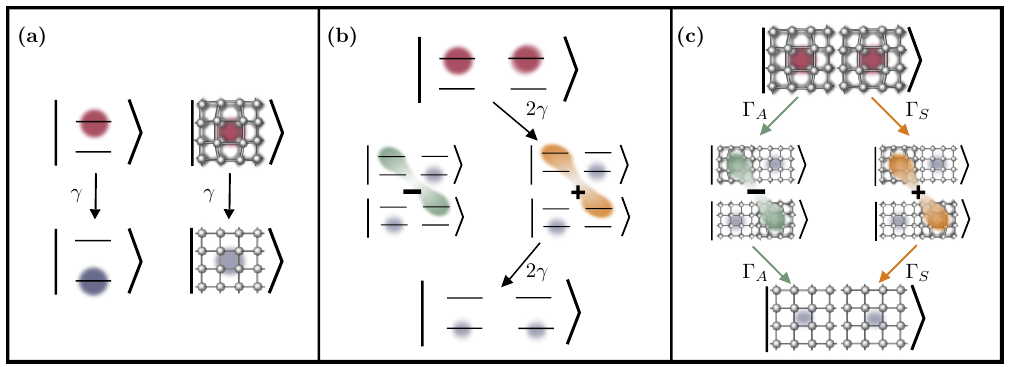}
    \caption{ (a) Single emitter case: For a two-level system the excited state (red) decays with rate $\gamma$ into the ground state (blue). The decay rate remains unchanged for an emitter embedded in a phonon environment (right). (b) Superradiance for two atomic emitters: Transitions towards the ground state (blue) happen with enhanced rate $2\gamma$ via the Dicke ladder involving the doubly excited state (red), the symmetric state (yellow), while the antisymmetric state (green) stays unoccupied. (c) Polaronic Dicke states of two condensed matter emitters: Optical transitions with rate $\Gamma_A$ involving the antisymmetric state (blue) are enabled, while the optical transition rate $\Gamma_S$ for the bright state (orange) is reduced compared to the phonon-free case.}\label{fig:figure_1}
\end{figure*}
\section{Phonon effects and radiative decay of a single emitter}\label{sec:TLS}
In this section, we determine the time evolution of a TLS simultaneously coupled to a structured phonon environment and a photon environment with a flat spectral density. We assume that the TLS is strongly coupled to the phonon environment, while the light-matter coupling is weak; owing to the large optical frequencies, the photon environment can be assumed to be at zero temperature.
While alternative and more involved treatments of the interplay between structured photon and phonon environments are available ~\cite{hughes2011, mitchison_non-additive_2018, maguire_environmental_2019,gribben_exact_2022, rouse_analytic_2022}, our present focus is on showing that a flat spectral density of the optical environment necessarily implies that the radiative decay rate remains unaffected by the coupling to phonons.
For this we first introduce the Hamiltonian of the TLS and its coupling to both environments, before transforming to the polaron frame and deriving a master equation for the radiative decay of the polaron. Afterwards, we obtain the time evolution of the electronic reduced density operator by applying the inverse polaron transform.
These calculations also serve as a vehicle for introducing the necessary analytical tools for obtaining analytic expressions in the two-emitter case.

\subsection{Single emitter Hamiltonian} 
We consider a single emitter that interacts with both its photon and phonon environment. The state of the whole system is then described by the density operator $\rho$ in its Hilbert space $\mathcal{H}_{S}\otimes\mathcal{H}_{PN}\otimes\mathcal{H}_{PT}$, where $\mathcal{H}_{S}$, $\mathcal{H}_{PN}$, and $\mathcal{H}_{PT}$ are the individual Hilbert spaces of the emitter, its phonon, and its photon environment, respectively.
The Hamiltonian can likewise be decomposed into three parts:
\begin{subequations}
\begin{align}
    H = H_{S} + H_{PN} + H_{PT}.
\end{align}
We model the emitter as a TLS with ground state $\ket{G}$ and excited state $\ket{X}$, and transition frequency $\omega_X$, while we set the ground state energy to zero.
Thus,
\begin{align}
    H_{S} &= \hbar\omega_X\sigma^+\sigma^- ,
\end{align}
where $\sigma^+ = \ket{X}\bra{G} = (\sigma^-)^\dagger$ are the usual Pauli operators. 
The coupling between the TLS and the electromagnetic vacuum is defined within the dipole and rotating wave approximation (RWA) \cite{cygorek_signatures_2023, christopher_c__gerry_introductory_2004} as
\begin{align}
    H_{PT} &= \hbar\sum_\textbf{k}\nu_\textbf{k}a^\dagger_\textbf{k}a_\textbf{k} + \hbar\sum_\textbf{k}g_\textbf{k}(a_\textbf{k}\sigma^+ + a_\textbf{k}^\dagger\sigma^-), 
\end{align}
where $a_\textbf{k}$ and $a^\dagger_\textbf{k}$ are the annihilation and creation operator of a photon in mode $\textbf{k}$ with energy $\hbar\nu_\textbf{k}$ and coupling strength $g_\textbf{k}$, respectively. 
Finally, the emitter is coupled to a condensed matter phonon environment. This environment can be constituted by the surrounding bulk material, in the case of GaAs QDs~\cite{krummheuer_pure_2005}, or vibrational modes of complex biological molecules \cite{clear_phonon-induced_2020}.
We denote the creation (annihilation) operator of a phonon in mode $\textbf{q}$ with energy $\hbar\omega_\textbf{q}$ by $b^\dagger_\textbf{q}$ ($b_\textbf{q}$). Then, the phonon contribution to the Hamiltonian reads \cite{nazir_modelling_2016,reiter_distinctive_2019} 
\begin{align}
    H_{PN}  &= \hbar\sum_\textbf{q}\omega_\textbf{q}b^\dagger_\textbf{q}b_\textbf{q} + \hbar\sigma^+\sigma^-\sum_\textbf{q}\xi_\textbf{q}(b_\textbf{q} + b_\textbf{q}^\dagger),
\end{align}
\end{subequations}
where $\xi_\textbf{q}$ denotes the coupling strength to mode $\textbf{q}$.
\subsection{Polaron transform}
Strong TLS-phonon coupling leads to the formation of a combined electronic and vibrational state, the polaron. 
Such electronic and vibrational mixing can be accounted for by transforming the Hamilton operator into the polaron frame (PF) \cite{hohenadler_lang-firsov_2007, nazir_modelling_2016}:
\begin{align}
\mathcal{U} &= \sigma^-\sigma^+ + \sigma^+\sigma^- \mathcal{D} \,, \\
\mathcal{D} &= \prod_\textbf{q}\exp\left((\xi_\textbf{q} b^\dagger_{\textbf{q}}-\left(\xi_\textbf{q}\right)^* b_{\textbf{q}})/ \omega_\textbf{q}\right)\,,
\end{align}
where $\mathcal{D}$ is the displacement operator.
Then, the transformed Hamilton operator $\tilde{H} :=  \mathcal{U}H\mathcal{U}^\dagger$, describes the dynamics of the polaron and its optical environment, characterised by the density operator $\tilde{\rho}$:
\begin{align}
    \tilde{H} =& \hbar(\omega_X-\omega_R)\sigma^+\sigma^- +\hbar\sum_{ \textbf{q}}\omega_\textbf{q}b_{\textbf{q}}^\dagger b_{\textbf{q}} + \hbar\sum_\textbf{k} \nu_\textbf{k} a_\textbf{k}^\dagger a_\textbf{k} \nonumber\\&+ \hbar\sum_\textbf{k}g_\textbf{k} (a_\textbf{k}^\dagger\mathcal{D}^\dagger\sigma^- + a_\textbf{k}\mathcal{D}\sigma^+)\,.\,\label{eq:PF_Hamilton_operator_TLS}
\end{align}
For the remainder of this article, we will follow the convention of indicating polaron frame operators with a tilde. 

By applying the polaron transform, one removes the TLS-phonon interaction term, leading to an energy shift of 
$\hbar\omega_R = \hbar\int_0^\infty d\omega J_{PN}(\omega)/\omega$
, the `reorganisation energy', where $J_{PN}(\omega) = \sum_\textbf{q}\xi_\textbf{q}^2\delta(\omega-\omega_\textbf{q})$ is the phonon spectral density. 
While being valid for all coupling strengths, this transformation is particularly useful if the phonon interaction is much stronger that the light-matter coupling. In this case one can derive a weak coupling master equation for the radiative decay of the polaron \cite{iles-smith_phonon_2017}.
\subsection{Radiative decay in the polaron frame}\label{sec:TLS_PME}
Having removed the potentially strong emitter-phonon interaction from the Hamiltonian by the use of the polaron transform, we proceed now by performing second-order perturbation theory for the light-matter interaction, and thus also the residual photon-mediated phonon coupling term in
 Eq.~\eqref{eq:PF_Hamilton_operator_TLS}. We derive the equations of motion of the reduced density operator for the electronic and vibrational degrees of freedom in the PF, $\tilde{\rho}_{S,PN} = \text{Tr}_{PT}[\tilde{\rho}]$, where we only trace out the photon environment. 
Moving into an interaction picture with respect to the free system and environment terms in Eq.~\eqref{eq:PF_Hamilton_operator_TLS}, $\tilde{H}_0 = \hbar(\omega_X-\omega_R)\sigma_+\sigma_- +\hbar\sum_{ \textbf{q}}\omega_\textbf{q}b_{\textbf{q}}^\dagger b_{\textbf{q}} + \hbar\sum_\textbf{k} \nu_\textbf{k} a_\textbf{k}^\dagger a_\textbf{k}$, we obtain the interaction Hamiltonian
\begin{align}
    H_I(t)    &=\hbar\sum_\textbf{k}g_\textbf{k} \left( \mathcal{D}^\dagger(t)\sigma^-a_\textbf{k}^\dagger e^{-i\omega_\textbf{k}t}+ \mathcal{D}(t)\sigma^+a_\textbf{k}e^{i\omega_\textbf{k}t}\right),
\end{align}
with $\omega_\textbf{k} = \omega_X-\omega_R-\nu_\textbf{k}$.
In accordance with the Born approximation, we assume that for each time step the density operator can be written as a direct product between the combined TLS and phonon part, and the electromagnetic field part, and that the photon environment is stationary, i.e.~$\tilde{\rho}(t) = \tilde{\rho}_{S, PN}(t)\otimes \rho_{PT}(0)$.
Then, the equation of motion for the reduced density operator in the polaron frame at time $t$ reads \cite{breuer_theory_2007}
\begin{align}
    \frac{d\tilde{\rho}_{S, PN}}{dt}= & -\frac{1}{\hbar^2}\int_0^tds\text{Tr}_{PT}\big\{[H_I(t), \nonumber\\&[H_I(t-s), \tilde{\rho}_{S, PN}(t-s)\otimes\rho_{PT}(0)]]\big\}.
\end{align}
Next we expand the double commutator and take the trace over the photon environment at zero temperature \footnote{The zero temperature assumption is justified for emitters in the optical range due to the high energy of optical photons. For different platforms, like superconducting qubits emitting in microwave region, the assumption of a zero temperature is not valid. However, the derivation presented here can be easily extended to include stimulated emission and absorption, leading to the same rate renormalizations for the absorption pathways.}, i.e.~without any initial photons present. We arrive at
\begin{widetext}
\begin{align}
\frac{d\tilde{\rho}_{S, PN}}{dt} = -\int_0^tds \sum_\textbf{k}|g_\textbf{k}|^2\big\{\sigma^+\sigma^-\mathcal{D}(t)\mathcal{D}^\dagger(t-s)\tilde{\rho}_{S, PN}(t-s)e^{i\omega_\textbf{k}s} - \sigma^-\mathcal{D}^\dagger(t-s)\tilde{\rho}_{S, PN}(t-s)\sigma^+\mathcal{D}(t)e^{i\omega_\textbf{k}s} \nonumber\\
- \sigma^-\mathcal{D}^\dagger(t)\tilde{\rho}_{S, PN}(t-s)\sigma^+\mathcal{D}(t-s)e^{-i\omega_\textbf{k}s}+ \tilde{\rho}_{S, PN}(t-s)\sigma^+\sigma^-\mathcal{D}(t-s)\mathcal{D}^\dagger(t)e^{-i\omega_\textbf{k}s}\big\}\,.\label{eq:TLS_Redfield_equation}
\end{align}
\end{widetext}
In a standard second-order perturbative treatment it is normally assumed that the environment correlation function decays quickly and one may therefore replace $\tilde{\rho}_{S, PN}(t-s) \approx \tilde{\rho}_{S, PN}(t)$ \cite{breuer_theory_2007}. This corresponds to the assumption of Markovianity. 
However, an infinitely short memory time also automatically follows from the assumption that the spectral density is flat, i.e.~$J_\text{PT}(\nu) = \sum_{\textbf{k}}|g_{\textbf{k}}|^2\delta(\nu-\nu_\textbf{k}) = (\gamma/2\pi)\theta(\nu)$, where $\theta(\nu)$ is the heaviside step function. Then,
\begin{align}
    &\sum_\textbf{k}|g_\textbf{k}|^2e^{\pm i\omega_ks} = \int_{-\infty}^\infty d\omega J_{PT}(\nu)e^{\pm i(\omega_X-\omega_R- \nu )s }\nonumber\\ 
    & = \frac{\gamma}{2\pi}\int_{-(\omega_X-\omega_R)}^\infty d\omega e^{\mp i\nu s }\approx \frac{\gamma}{2\pi}\int_{-\infty}^\infty d\nu e^{\mp i\nu s } = \gamma\delta(s),\label{eq:flat_density_approximation}
\end{align}
where the lower integration bound can be extended to negative infinity due to the large optical frequencies involved.
Inserting this into Eq.~\eqref{eq:TLS_Redfield_equation}, we obtain
\begin{align}
    \frac{d\tilde{\rho}_{S, PN}}{dt}(t) = \gamma\big\{\sigma^-\mathcal{D}^\dagger(t)&\tilde{\rho}_{S, PN}(t)\sigma^+\mathcal{D}(t) \nonumber\\&- \frac{1}{2}[\sigma^+\sigma^-, \tilde{\rho}_{S, PN}(t)]_+\big\}.
\end{align}

In the polaron frame the TLS-phonon interaction has been absorbed into the definition of the polaronic excitation.
This allows for a Born approximation for the residual phonon degrees of freedom. Following standard polaron theory, we obtain a Lindblad propagator for the polaron by substituting phonon operators by their expectation values with respect to a thermal state \cite{nazir_modelling_2016, rouse_optimal_2019,clear_phonon-induced_2020, rouse_analytic_2022}. For this, we replace
\begin{align}
    \mathcal{D}^\dagger(t)\tilde{\rho}_{S, PN}(t)\mathcal{D}(t) \to\,&\tilde{\rho}_{S, PN}(t)\text{Tr}[\mathcal{D}^\dagger(t){\rho}_{G}\mathcal{D}(t)]\nonumber\\& = \tilde{\rho}_{S, PN}(t),
\end{align}
where ${\rho}_{G}$ is the density operator of a Gibbs state \cite{breuer_theory_2007}.
This results in
\begin{align}
    \frac{d\tilde{\rho}_{S, PN}}{dt}(t) = \gamma\mathcal{L}[\sigma_-]\tilde{\rho}_{S, PN}(t)\,\label{eq:TLS_ME_PF},
\end{align}
 where the Lindblad dissipator \cite{lindblad_generators_1976} for a given operator $\hat{O}$ acting on a density operator $\rho$ reads $\mathcal{L}[\hat{O}]\rho = \hat{O}\rho\hat{O}^\dagger - (\left( \hat{O}^\dagger\hat{O}\rho - \rho\hat{O}^\dagger\hat{O} \right)/2$.
Equation \eqref{eq:TLS_ME_PF} shows that a TLS excitation decays with rate $\gamma$, which is only determined by the light-matter coupling. 
Previously, it has been argued, that the transition rate has to be renormalised due to a Franck-Condon displacement of the excited state \cite{roy_microscopic_2010, hughes2011}, while others reached the conclusion that this is not the case \cite{nazir_modelling_2016}. 
In the former case, the rate would change corresponding to the renormalization of the transition dipole. 

This seeming contradiction arises from assuming different hierarchies of correlation times for the phonon and the photon environment. Making an explicit Markov assumption (negligibly short correlation time) for the photon environment by choosing a flat spectral density we disambiguate this hierarchy and obtain the same results as first taking the Markov limit for the photon environment \cite{iles-smith_phonon_2017, rouse_analytic_2022}.Typically, the phonon correlation time exceeds the photon correlation time significantly, except in the case of strongly coupled high quality microcavities
\footnote{Note that the opposite typically is true when comparing the time scale of decoherence and decay. In most cases the radiative lifetime is significantly longer than the dynamics induced by the phonon coupling. This justifies the perturbative and strong coupling treatment, respectively.}.

Note that whilst the overall decay rate is not altered by the phonon coupling, it includes emission via the zero-phonon line as well as the phonon sidebands. 
The fraction that is emitted via the zero-phonon line does, in fact, depend on the strength of the emitter-phonon coupling and on the temperature [See Ref.~\cite{iles-smith_phonon_2017}].
\subsection{Dynamics of the bare electronic excitation}\label{sec:TLS_lab_frame}
In the previous section, we found the equations of motion governing the decay of a polaronic TLS excitation. In this section, we obtain the time evolution of a TLS coupled to a phonon environment that is initialized in an arbitrary electronic state in the original, i.e. not polaron transformed, frame, while the phonon environment is in a thermal state described by the density operator $\rho_{PN}(0)$.

For this, we transform the initial state into the PF, apply the dynamical map $\mathcal{M}(t):\tilde{\rho}_{S, PN}(0) \mapsto \tilde{\rho}_{S, PN}(t)$, mapping the initial state represented in the polaron frame to its state at time $t$, and transform the result back by applying the inverse polaron transform.
Consequently, the time evolution of the density operator for the bare electronic state is given by 
\begin{align}
    {\rho}_S(t) =&\text{Tr}_{PN}\left[\mathcal{T}^{-1}(t)\left\{\mathcal{M}(t)\mathcal{T}[{\rho}_S(0)\otimes\rho_{PN}(0)]\right\}\right] ~,\label{eq:time_evolution_density_matrix_TLS}
\end{align}
where $\mathcal{T}[\rho] = \mathcal{U}{\rho}\mathcal{U}^\dagger$ and $\mathcal{T}^{-1}(t)[\tilde{\rho}] = \mathcal{U}^\dagger(t){\tilde{\rho}}\mathcal{U}(t)$ are the polaron transform and inverse polaron transform, where $\mathcal{U}(t) =  e^{\frac{itH_E}{\hbar}}\mathcal{U}e^{-\frac{itH_E}{\hbar}}$, with $H_{E} = \hbar\sum_\textbf{q}\omega_\textbf{q}b^\dagger_\textbf{q}b_\textbf{q}$. The time dependence of the inverse polaron transform stems from moving back from the interaction picture to the Schrödinger picture.

The dynamical map $\mathcal{M}(t)$ can be obtained by solving the equations of motion given by Eq.~\eqref{eq:TLS_ME_PF} for arbitrary initial states $\rho_S(0)$. Setting the polaron energy to zero, i.e.~$\omega_X-\omega_R = 0$, we obtain
\begin{subequations}
\begin{align}
    \tilde{\rho}_{X, X}(t) &= \tilde{\rho}_{X, X}(0)e^{-\gamma t}\,,\\
    \tilde{\rho}_{G, G}(t) &= \tilde{\rho}_{G, G}(0)-\tilde{\rho}_{X, X}(0)(e^{-\gamma t}-1)\,,\\
    \tilde{\rho}_{G, X}(t) &= \tilde{\rho}_{G, X}(0)e^{-\gamma t/2}\,,
\end{align}
where we have defined $\tilde{\rho}_{\alpha, \alpha^\prime}(t) = \bra{\alpha}\tilde{\rho}_{S, PN}(t)\ket{\alpha^\prime}$ with $\alpha,\alpha^\prime  \in \{G, X\}$.
\end{subequations}
Thus, for an arbitrary initial bare electronic density operator with elements ${\rho}_{\alpha, \alpha^\prime}(t) = \bra{\alpha}{\rho}_{S}(t)\ket{\alpha^\prime}$, we obtain
\begin{align}
    \rho_S(t) =& \rho_{X, X}(0)e^{-\gamma t}\ket{X}\bra{X} \nonumber\\
    &+ {\rho}_{G, X}(0)\langle\mathcal{D}(t)\mathcal{D}^\dagger\rangle e^{-\gamma t/2}\ket{G}\bra{X} \nonumber\\
    &+ {\rho}_{X, G}(0)\langle\mathcal{D}^\dagger(t)\mathcal{D}\rangle e^{-\gamma t/2}\ket{X}\bra{G} \nonumber\\
     &+ (1-(1-{\rho}_{G, G}(0))e^{-\gamma t})\ket{G}\bra{G} .\label{eq:TLS_time_ev}
\end{align}
We explicitly calculate the expression $\langle\mathcal{D}(t)\mathcal{D}^\dagger\rangle$ in Sec. S3.3.~of the supplementary material (SM)~\cite{supplement} and obtain
\begin{align}
    \langle\mathcal{D}^\dagger(t)\mathcal{D}\rangle = (\langle\mathcal{D}^\dagger\mathcal{D}(t)\rangle)^* = \kappa^2\exp(\Phi(t)).
\end{align}
Here, we have introduced the usual environment correlation function 
\begin{align}
    \Phi(t) = \int_{-\infty}^{\infty} \frac{J_{PN}(\omega)}{\omega^2}\left(\coth(\beta\omega)\cos(\omega t) - i\sin(\omega t)\right)\,,
\end{align}
and the expectation value of the displacement $\kappa = \exp(\Phi(0)/2)$.

For concreteness, we specify the phonon coupling by using a standard spectral density for self-assembled GaAs QDs \cite{koong_coherent_2021, cygorek_simulation_2022, wiercinski_phonon_2023}, which is strongly dominated by the deformation potential coupling to longitudinal-acoustic phonons and can be derived from microscopic theory \cite{krummheuer_theory_2002,mahan_many-particle_2000}. The corresponding coupling spectral density is
\begin{align}
    J_{PN}(\omega) &= \sum_\textbf{q}\xi_\textbf{q}^2\delta(\omega-\omega_\textbf{q}) \nonumber\\
    &= \frac{\omega^3}{2\mu\hbar c_s^5}\left(D_e e^{-\frac{\omega^2}{\omega_e^2}}-D_h e^{-\frac{\omega^2}{\omega_h^2}}\right)^2\,,\label{eq:SD_QD}
\end{align} 
with the mass density $\mu$, the electron (hole) deformation potential $D_e$ ($D_h$), the electron (hole) cutoff frequency $\omega_e$ ($\omega_h$) and the speed of sound $c_s$. We use realistic values for InGaAs/GaAs QDs at a temperature of $T=4$~K and a radius of approximately $4$~nm (cf.~Ref.~\cite{krummheuer_pure_2005}).  Numerical parameter values are given in Sec.~S4 of the SM \cite{supplement}.

\begin{figure}
    \includegraphics[width=\linewidth]{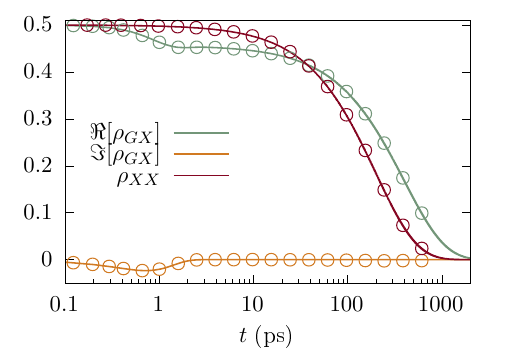}
    \caption{Dynamics of a two-level system initialized in the coherent superposition $(\ket{X}+\ket{G})/2$ coupled to a phonon environment at $T=4$~K as well as a photon environment with flat spectral density corresponding to a radiative lifetime of $\gamma^{-1} = 0.2$~ns. We find a perfect agreement of our analytic results (solid lines), which predict no change in radiative decay rate with our numerically exact simulations (circles). A logarithmic time axis has been used to cover the two time scales of the system evolution.}\label{fig:TLSDynamics}
\end{figure}
Equations \eqref{eq:TLS_ME_PF} and \eqref{eq:TLS_time_ev} predict that, under the assumption of a flat optical spectral density Eq.~\eqref{eq:flat_density_approximation}, radiative decay is not altered by the coupling to the phonon environment. This prediction is corroborated by numerically exact calculations that simultaneously include both environments [cf.~Fig.~\ref{fig:TLSDynamics}]. For this we use a state-of-the-art process tensor method that delivers numerically converged system dynamics irrespective of the system-environment coupling strength. This method also takes into account possible cross-interactions of different environments by efficiently combining their process tensors \cite{cygorek_simulation_2022, cygorek_sublinear_2023}. For our simulations we use the ACE code described in Ref.~\cite{cygorek_ace_2024}.

Further, looking at the coherences between ground and excited state, we observe, both in our analytical as well as in our numerically exact results, an initial,  non-complete, drop known from the independent boson model which arises due to the super-Ohmic nature of the chosen coupling spectral density \cite{nazir_modelling_2016}. Superimposed is a long-time decay of the coherences due to the electromagnetic coupling. 
\section{Phonon influence on superradiant decay of two TLSs}\label{sec:2TLS}
Compared to emission of a single dipole, the radiative decay of two or more emitters is more complex due to the impact of inter-emitter coherences on the radiation process. 
Most notably, in the case of superradiance, transitions of $N$ emitters along a collective symmetric superposition state lead to an transient enhancement of the radiative decay rate that is proportional to $N^2$. At the same time, all non-symmetric states decouple from the electromagnetic environment  \cite{gross_superradiance_1982}. 

In this article we focus on only two emitters. In this case the presence of a phonon environment enables transitions between the bright symmetric and the decoupled antisymmetric electronic superposition state. Previous works have shown that these processes can be represented in a polaron picture by the decay rate of the symmetric (antisymmetric) polaronic state being reduced (enhanced) with respect to the phonon-free case \cite{rouse_optimal_2019, tiranov_collective_2023}. In this section we recapitulate the necessary steps to arrive at this rate renormalisation. However, it will become apparent in Sec.~\ref{sec:2TLS_excitation_dependence} that this is insufficient to describe the evolution of bare electronic state.
Therefore, the second goal of this section is to show how the polaron description can be extended to analytically derive the time evolution of bare electronic emitter states.
\subsection{Bare system Hamiltonian}
In analogy to the previous Sec.~\ref{sec:TLS} we introduce two emitters (QDs) coupled to individual phonon environments and a common electromagnetic environment. The corresponding Hamiltonian is
\begin{align}
    H = H_{S} + H_{PT} + H_{PN}\,,\label{eq:2TLS_Hamiltonian}
\end{align}
where $H_{S}$ is the Hamiltonian of the two emitters and $H_{PN}$ describes two phonon environments and their interaction with their respective emitters.  $H_{PT}$ describes a common electromagnetic environment and its interaction with both emitters.

We describe the two emitters as TLSs, with ground states $\ket{G_j}$ and excited states $\ket{X_j}$, $j=1,2$, and denote the corresponding raising and lowering operators by $\sigma^+_{1/2} = \ket{X_{1/2}}\bra{G_{1/2}}$ and $\sigma^-_{1/2} = (\sigma^+_{1/2})^\dagger$, respectively. 
We assume that the two TLSs possess transition frequencies $\omega_1$ and $\omega_2$, i.e.,
\begin{align}
    H_\text{S} = \hbar\sum_{j=1,2} \omega_j \sigma^+_j\sigma^-_j\,.
\end{align}
In general, the two emitters couple with different phases to the electromagnetic field modes, according to their position, $\textbf{R}_1$ and $\textbf{R}_1$ \cite{cygorek_signatures_2023}.
However, when their spatial separation is sufficiently below the wavelength of the emitted light, i.e., $(\textbf{R}_1-\textbf{R}_2)\cdot\textbf{k}\approx0$ for all relevant wave vectors $\textbf{k}$, then the light-matter Hamiltonian can be written as
\begin{align}
    H_\text{PT} =& \hbar\sum_\textbf{k} \nu_\textbf{k} a_\textbf{k}^\dagger a_\textbf{k} + \hbar\sqrt{2}\sum_\textbf{k}g_\textbf{k} (\sigma_S^-a_\textbf{k}^\dagger + \sigma_S^+a_\textbf{k})\,,
\end{align}
with $\sigma_{S/A}^+ =(\sigma^+_1 \pm \sigma^+_2)/\sqrt{2}$. Thus, we introduce the Dicke basis states of the single excitation manifold,
\begin{subequations}
\begin{align}
    \ket{\Psi_S} &= \sigma_S^+\ket{G_1, G_2} = \frac{1}{\sqrt{2}}\left(\ket{X_1,G_2} +\ket{G_1,X_2} \right)\label{eq:coll_basis_S}, \\
    \ket{\Psi_A} &= \sigma_A^+\ket{G_1, G_2} = \frac{1}{\sqrt{2}}\left(\ket{X_1,G_2} - \ket{G_1,X_2} \right)\label{eq:coll_basis_A}
\end{align}
\end{subequations}
which we refer to as symmetric and antisymmetric state, respectively. The antisymmetric state $\ket{\Psi_A}$ is completely decoupled from the electromagnetic field, while the electromagnetic field coupling strength to the symmetric state $\ket{\Psi_S}$ is enhanced by a factor of $\sqrt{2}$ compared to the individual emitter coupling. Hence, transitions through the bright symmetric state experience an increase in the radiative decay rate by a factor of two, whereas the anti\-symmetric state remains dark [cf.~Fig.~\ref{fig:figure_1}~(b)].
\subsection{Polaron transform and radiative decay rate renormalisation}\label{sec:2TLS_lab_frame} 
We now introduce the vibrational environment, which, for self-assembled QDs, is constituted by LA phonons of the bulk material \cite{krummheuer_pure_2005}. In this three-dimensional environment phonon wave packets disperse quickly, limiting the phonon coherence length to a few nanometers \cite{mccutcheon_separation-dependent_2010,wigger_energy_2014}. Due to the distance between two QDs typically being much larger than this, the vibrational environment can be assumed as a local phonon bath for each individual emitter \cite{wiercinski_phonon_2023}:
\begin{align}
    H_\text{PN} = \hbar\sum_{j,\textbf{q}} \left[\omega_\textbf{q}b_{\textbf{q}, j}^\dagger b_{\textbf{q},j} + \sigma_j^+\sigma_j^-\xi_{\textbf{q},j}(b_{\textbf{q}, j}^\dagger + b_{\textbf{q},j} )\right]
\end{align}
with creation (annihilation) operators of the $\textbf{q}$-mode of the $j$-th phonon environment $b_{\textbf{q}, j}^\dagger$ ($b_{\textbf{q}, j}$), mode frequencies $\omega_\textbf{q}$, and phonon coupling strengths $\xi_{\textbf{q},1} =  \xi_{\textbf{q},2}$, which we assume to be real \footnote{Note that for QDs it is possible to fulfil both requirements, i.e. being far apart enough for the phonon environments being independent and close enough for being superradiant, due to the large discrepancy between the wave lengths of phonons and optical photons. Also, other condensed matter systems like molecular dimers are typically modelled this way.}.

Similar to the case of a single TLS, the dynamics of the polaronic excitations, captured in the PF, will be described using a Markovian master equation, as the dominant non-Markovian environment effects have been absorbed into the definition of this vibronic excitation. For two phonon environments, the polaron transformation becomes
\begin{align}
\mathcal{U}_j &= \sigma^-_j\sigma^+_j + \sigma^+_j\sigma^-_j D_j\,
\end{align}
with the displacement operator for an indivdual QD $j$
\begin{align}
    D_{j} &= \prod_\textbf{q}\exp\left((\xi_{\textbf{q},j} b^\dagger_{\textbf{q}, j}-\xi_{\textbf{q},{j}}^* b_{\textbf{q}, j})/ \omega_\textbf{q}\right).
\end{align}
Then, the successive application of polaron transform operators can be expressed in the collective basis:
\begin{align}
    \mathcal{U}_1\mathcal{U}_2 =& \ket{G_1, G_2}\bra{G_1, G_2} + \ket{X_1, X_2}\bra{X_1, X_2}D_1D_2\nonumber\\
    &+ (\ket{\Psi_S}\bra{\Psi_A} + \ket{\Psi_A}\bra{\Psi_S})\mathcal{D}_- \nonumber\\&+ (\ket{\Psi_S}\bra{\Psi_S} + \ket{\Psi_A}\bra{\Psi_A})\mathcal{D}_+
\end{align}
with $\mathcal{D}_{\pm} = \left(D_1 \pm D_2\right)/2$. 
We obtain the time evolution of the reduced density operator for the bare electronic states by generalizing the approach from Sec.~\ref{sec:TLS} to two emitters. This means solving 
\begin{align}
    {\rho}_S(t) =&\text{Tr}_{PN}\left[\mathcal{T}^{-1}(t)\left\{\mathcal{M}(t)\mathcal{T}[{\rho}_S(0)\otimes\rho_{PN}(0)]\right\}\right]\label{eq:time_evolution_density_matrix},
\end{align}
where
\begin{align}
    \mathcal{T}[\rho] &= \mathcal{U}_1\mathcal{U}_2{\rho}\mathcal{U}_2^\dagger\mathcal{U}_1^\dagger\,, \\
    \mathcal{T}^{-1}(t)[\rho] &= e^{\frac{itH_E}{\hbar}}\mathcal{T}^{-1}[\tilde{\rho}]e^{-\frac{itH_E}{\hbar}}\nonumber\\ 
    &= \mathcal{U}_1^\dagger(t)\mathcal{U}_2^\dagger(t){\tilde{\rho}}\mathcal{U}_2(t)\mathcal{U}_1(t)\label{eq:2TLS_polaron_transform}
\end{align}
with $\mathcal{T}$ ($\mathcal{T}^{(-1)}$) being the (inverse) polaron transform superoperators for both phonon environments, and where $\mathcal{U}_{1/2}(t) = e^{\frac{itH_E}{\hbar}}\mathcal{U}_{1/2} e^{-\frac{itH_E}{\hbar}}$. 
The undisturbed phonon Hamiltonian reads $H_\text{E} = \hbar\sum_{j, \textbf{q}}\omega_\textbf{q}b_{\textbf{q},j}^\dagger b_{\textbf{q},j}$.
In analogy to Sec.~\ref{sec:TLS}, $\mathcal{M}(t)$, describes the dynamical map for the combined system and phonon environment degrees of freedom.

The impact of different dipole transition energies and a non-flat spectral density on superradiance has been discussed elsewhere  \cite{PhysRevB.90.125307, mccauley_accurate_2020, PhysRevA.105.062207}, albeit not with an additional phonon coupling. For this article, we restrict ourselves to the special case of a flat spectral density for the electromagnetic environment and identical emitters.  
This means that the phonon environments, as well as the dipole transition energies $\omega_{1/2} = \omega_X$, are identical. While we provide a more extensive derivation in Sec.~S3 of the SM \cite{supplement}, here we only recapitulate its essential steps. First, we apply the polaron transforms $\mathcal{U}_j$, to the interaction picture Hamiltonian \eqref{eq:2TLS_Hamiltonian}, which results in
\begin{align}
    \tilde{H} 
    =& \hbar\sum_j\sigma^+_j\sigma^-_j(\omega_j-\omega^R_j) + \hbar\sum_\textbf{k} \nu_\textbf{k} a_\textbf{k}^\dagger a_\textbf{k} \nonumber\\&+ \hbar\sum_{j,\textbf{k}}g_\textbf{k} \left(a_\textbf{k}^{\dagger}{D}_j^\dagger(t)\sigma^-_j + a_\textbf{k}{D}_j(t)\sigma^+_j\right)\,\label{eq:PF_Hamilton_operator_SR}
\end{align}
with individual emitter renormalization energies $\omega^R_{j} = \sum_\textbf{q}|\xi_{\textbf{q},j}|^2/\omega_\textbf{q}$.
Analogously to Sec.~\ref{sec:TLS}, we treat the light-matter coupling perturbatively to second-order and obtain a master equation for the combined phonon and electronic degrees of freedom
\begin{align}
\frac{d\tilde{\rho}_{S, PN}}{dt} = \gamma \sum_{i, j = 1}^{2}&\big\{\big({D}_i^\dagger\sigma^-_i\big)\tilde{\rho}_{S, PN}\big({D}_j\sigma^+_j\big)\nonumber\\
&- \frac{1}{2}\big[\big({D}_i\sigma^+_i\big)\big({D}_j^\dagger\sigma^-_j\big), \tilde{\rho}_{S, PN}\big]_+\big\}.
\end{align}

To arrive at a reduced master equation in Lindblad form \cite{breuer_theory_2007} we again follow the standard polaron master equation treatment by replacing phonon operators by their expectation value with respect to a Gibbs state $\tilde{\rho}_{PN} = \rho_{G, 1} \otimes \rho_{G, 2}$. With cross-interactions between the polaronic Dicke states dropping out due to the phonon environments being identical, this yields a Lindblad propagator for the reduced system density matrix in the PF [cf.~SM~Sec.~S2~\cite{supplement}]
\begin{align}
    \frac{d\tilde{\rho}_{S, PN}}{dt} = \Gamma_S \mathcal{L}[\sigma_S]\tilde{\rho}_{S, PN} + \Gamma_A \mathcal{L}[\sigma_A]\tilde{\rho}_{S, PN}\label{eq:PF_ME}
\end{align}
with $\Gamma_{S/A} = (1\pm\kappa^2)\gamma$, and $\kappa = \textrm{Tr}[D_{1/2}\rho_{PN, 1/2}(0)]$. For $\kappa = 1$, we recover the phonon-free case, while $\kappa = 0$ leads to independent emission.
From the polaron frame equations of motion Eq.~\eqref{eq:PF_ME} [cf.~SM~Sec.~\cite{supplement}] the propagator $\mathcal{M}(t)$ [cf.~Eq.~\eqref{eq:time_evolution_density_matrix}] can be determined.

Therefore, in the PF, we find that due to the electron-phonon coupling the decay rate of the symmetric polaronic state $\ket{\Psi_S}$ is reduced compared to the phonon free case, i.e., $\Gamma_S < 2\gamma$. Meanwhile, the antisymmetric polaronic state $\ket{\Psi_A}$ decays with a nonzero rate $\Gamma_A >0$. This is depicted in Fig.~\ref{fig:figure_1}~(c). Note, that these decay rates are a limiting case of those previously derived in Ref.~\cite{rouse_optimal_2019} for molecular dimers. 
In the following section, we make the case that these decay rates do not tell the full story: The way how the two-emitter system is excited  can lead to mixing between the symmetric and antisymmetric polaronic Dicke state. This mixture in turn determines the time evolution of the system not only on short, but also on extended time scales.

\section{Dynamics of two superradiant emitters in the presence of phonons}\label{sec:results}
\subsection{Phonon-induced decoherence of two emitters}\label{sec:two_TLS_decoherence}
In this section, we determine the time evolution of occupations of the bare electronic Dicke states $\ket{\Psi_S}$ and $\ket{\Psi_A}$ resulting from the coupling to the phonon environment.
In Ref.~\cite{wiercinski_phonon_2023}, we argued that the phonon influence within the single-excitation manifold can be understood by mapping the single-excitation manifold to an independent boson model. In this section, we 
provide more insight by analytically deriving the dynamics induced by the phonon coupling by utilising the framework presented in Sec.~\ref{sec:2TLS_lab_frame}.

Within this framework, the case of free decoherence corresponds to zero coupling strength to the photon environment, i.e.~$\mathcal{M}(t) = 1$. Therefore, Eq.~\eqref{eq:time_evolution_density_matrix} simplifies to applying the polaron transform at time zero and the back transform at time $t$.

In line with our assumption that the two emitters are identical and couple identically to the electromagnetic field and their phonon environments, we only consider states which are symmetric or antisymmetric with respect to exchange of particle. Thus for any electronic state $\rho_S$ it must hold that 
\begin{align}
     \text{Tr}[\bra{\Psi_A}\rho_S\ket{\Psi_S}] 
     = \frac{1}{2}(&\bra{X_1, G_2}\rho_S\ket{G_1, X_2}\nonumber\\&-\bra{G_1, X_2}\rho_S\ket{X_1, G_2}) = 0
\end{align}
and we can write the initial state as ${\rho}_S(0) = \rho_{SS}(0)\ket{\Psi_S}\bra{\Psi_S} + \rho_{AA}(0)\ket{\Psi_A}\bra{\Psi_A}$. We obtain:
\begin{align}
    \rho_S&(t) =a_-(t)(\rho_{SS}(0)\ket{\Psi_A}\bra{\Psi_A} + \rho_{AA}(0)\ket{\Psi_S}\bra{\Psi_S}) \nonumber\\
    &+ a_+(t)(\rho_{SS}(0)\ket{\Psi_S}\bra{\Psi_S} + \rho_{AA}(0)\ket{\Psi_A}\bra{\Psi_A})
\end{align}
with
\begin{align}
    a_+(t) &= \sum_{s_1, s_2}\langle\mathcal{D}_{s_1}^\dagger(0)\mathcal{D}_{s_1}(t)\mathcal{D}_{s_2}^\dagger(t)\mathcal{D}_{s_2}(0)\rangle \nonumber\\
    &= \frac{1}{2}\left(1+\kappa^4\exp(2\Re[\Phi(t)])\right)\nonumber,\\
    a_-(t) &= \sum_{s_1, s_2}\langle\mathcal{D}_{s_1}^\dagger(0)\mathcal{D}_{-s_1}(t)\mathcal{D}_{s_2}^\dagger(t)\mathcal{D}_{-s_2}(0)\rangle \nonumber\\
    &= \frac{1}{2}\left(1-\kappa^4\exp(2\Re[\Phi(t)])\right)\nonumber
\end{align}
for $s_1, s_2 \in \{+,-\}$.

\begin{figure}
    \centering
    \includegraphics[width=\linewidth]{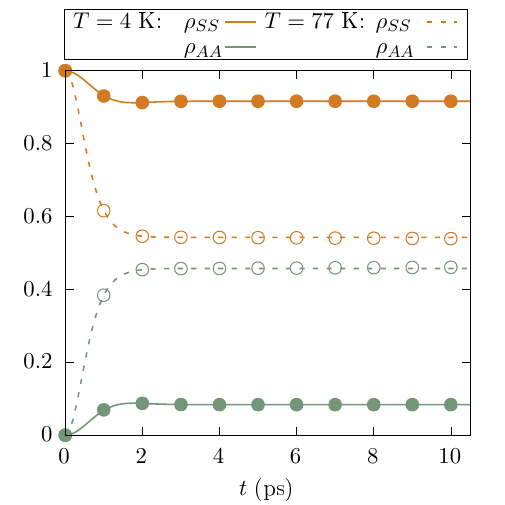}
    \caption{Decoherence behaviour of two dipoles in the absence of radiative decay, if initialized in the symmetric state $\ket{\Psi_S}$. Depicted are the symmetric (orange line) and antisymmetric (green line) state occupations for $T = 4$~K and $T = 77$~K. Additionally, we show numerically exact simulations for $T=4$~K (solid circles) and $T=77$~K (rings).}\label{fig:AnalyticalDecoherence}
\end{figure}

For initialization of the system with $\rho_{SS}(0) = 1$, the occupations for the symmetric and the antisymmetric state are shown in Fig.~\ref{fig:AnalyticalDecoherence} for temperatures of $T=4$~K and $T=77$~K. As one can see, the two phonon environments lead to partial decoherence during the formation of the polaron resulting in transitions from the initial antisymmetric state to the symmetric state. The amount of mixing increases with temperature. To gain additional insight, we perform a change of basis into the individual emitter basis and find:
\begin{align}
    \rho_S(t) &=\rho_{11}(0)\ket{X_1, G_2}\bra{X_1, G_2} + \rho_{22}(0)\ket{G_1, X_2}\bra{G_1, X_2} \nonumber\\
    &+ \rho_{12}(0)\left(\kappa^2e^{\Re[\Phi(t)]}\right)^2\ket{G_1, X_2}\bra{X_1, G_2} \nonumber\\
    &+ \rho_{21}(0)\left(\kappa^2e^{\Re[\Phi(t)]}\right)^2\ket{X_1, G_2}\bra{G_1, X_2}\,,
\end{align}
where $\rho_{12}(0) = \rho_{21}(0)= ( \rho_{SS}(0) - \rho_{AA}(0))/2$ are the coherences between the QDs, and $\rho_{11}(0) = \rho_{22}(0)=(\rho_{SS}(0) + \rho_{AA}(0))/2$ their individual occupations. Comparing this to results for an individual QD given by Eq.~\eqref{eq:TLS_time_ev} we find 
that two phonon environments act on the inter-emitter coherences of two TLSs like one environment acting on one TLS with the new correlation function $\Phi(t) + \Phi^*(t)$. 
This corresponds to the aforementioned mapping of the single excitation manifold to the independent boson model \cite{wiercinski_phonon_2023}.
Due to the super-Ohmic nature of the quantum dot spectral density, the mixing between the dark and the bright state is not complete, reflecting the incomplete decay of inter-emitter coherences. For other types of spectral densities, most notably Ohmic ones, which are relevant for many biological systems \cite{knox_low-temperature_2002, caycedo-soler_exact_2022, somoza_driving_2023}, this mixing can be stronger and lead to a complete decay of coherences. However, note that (sub-)Ohmic spectral densities lead to divergences in the polaron approach.

\subsection{Weak driving of superradiant QDs}\label{sec:2TLS_weak_excitation}
Experimentally, Dicke states can be excited by utilizing pulsed laser driving of both QDs \cite{tiranov_collective_2023}. In this section we show that the laser pulse duration determines if Dicke states of electronic or polaronic nature are excited and that their different dynamics are accentuated in the weak driving limit.
We calculate the time dependence of the mean excitation number $n(t) = \langle \sigma_S^+\sigma_S^- + \sigma_A^+\sigma_A^-\rangle$, from which the emitted intensity can be obtained by differentiation, $I(t) = -dn(t)/d t$.
First, we show that for short pulses the long-time evolution can approximately be obtained analytically by solving Eq.~\eqref{eq:time_evolution_density_matrix}. Then, we investigate how the lifetime, obtained by a monoexponential approximation, changes for different initial states. Afterwards, we compare these results to the time evolution for longer pulse durations, which can be captured by solving the PF equations of motion.

Like the common electromagnetic environment in Sec.~\ref{sec:2TLS_lab_frame}, the exciting laser couples via the symmetric state operator to the two-TLS system, thus, naturally, it couples to the bare electronic symmetric transitions. We describe this using the Rabi Hamilton operator
\begin{align}
    H_\text{Pulse}(t) = \frac{\Omega(t)}{2}(\sigma^+_S+ \sigma^-_S)\,
\end{align}
with a Gaussian-shaped envelope 
\begin{align}
    \Omega(t) = \frac{A}{\sqrt{2\pi}\sigma}e^{-\frac{(t-t_c)^2}{2\sigma^2}}
\end{align}
with pulse area $A$, which allows to control the overall excitation put into the system. While varying the pulse length $\sigma = \tau_\mathrm{FWHM}/\sqrt{8\ln(2)}$, we adjust for the pulse position to be $t_c = 3\sigma$ to account for more than $99.5\%$ of the pulse area to be captured in our simulations.
If the pulse area is small, the doubly excited state remains approximately unoccupied, as its excitation is a higher order process that requires transitioning through the single excitation manifold. In the following section we use a pulse area of $A = \pi/8$, amounting to a doubly excited state occupation of less than $1\%$ of the single-excitation manifold occupation.  Therefore, we can restrict our analysis to the single-excitation manifold.
\subsubsection{Excitation with short pulses}\label{sec:2TLS_short_pulse}
\begin{figure}
    \centering
    \includegraphics[width=\linewidth]{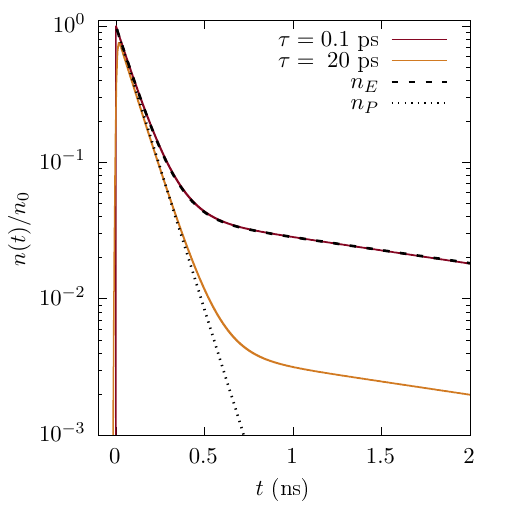}
    \caption{Normalized excitation number of two superradiant emitters driven by a laser pulse at $t = 0$ with area $A=\pi/8$ for two different pulse lengths. Normalization factors: $n_0 = 0.036$ for $\tau = 0.1$~ps, and $n_0 = 0.027$ for $\tau = 20$~ps. We show, that these two cases are well-represented by the mean excitation number of electronic excitations, $n_E$, as calculated via Eq.~\eqref{eq:time_evolution_density_matrix} with ${\rho}_S(0) = \ket{\Psi_S}\bra{\Psi_S}$, and the excitation number ${n}_P$, obtained by solving the PF master equation with $\tilde{\rho}_S(0) = \ket{\Psi_S}\bra{\Psi_S}$, respectively.}\label{fig:PulseLengthDependenceNumber}
\end{figure}

If the pulse duration is very small---smaller than the dynamics induced by the phonon coupling---the laser pulse effectively acts as a $\delta$-like excitation, and the situation corresponds to an initial value problem as described in Sec.~\ref{sec:2TLS_lab_frame} with ${\rho}_S(0) = \ket{\Psi_S}\bra{\Psi_S}$. We show that the normalized mean excitation number can be obtained by solving Eq.~\eqref{eq:time_evolution_density_matrix} for $\mathcal{M}(t) = \exp[(\Gamma_S\mathcal{L}_{\sigma^-_S}+\Gamma_A\mathcal{L}_{\sigma^-_A})t]$ and ${\rho}_S(0) = \ket{\Psi_S}\bra{\Psi_S}$. We compare this prediction with numerically exact results [cf.~SM~Sec.~4~\cite{supplement}], in which we explicitly account for the pulsed excitation in Fig.~\ref{fig:PulseLengthDependenceNumber}.

While we show the whole derivation of our analytical result in the SM [cf.~Sec.~S3.6~\cite{supplement}] we here only state the final result: We find the occupations of the symmetric and antisymmetric state to evolve according to
\begin{align}
    \bra{\Psi_S}{\rho}_{S}(t)\ket{\Psi_S} =&e^{-\Gamma_St}\mathcal{E}_{++}(t) + e^{-\Gamma_At}\mathcal{E}_{-+}(t)-e^{- \gamma t}\mathcal{E}_0(t)        \label{eq:rhoSS_single_excitation_manifold}\,,\\
       \bra{\Psi_A}{\rho}_{S}(t)\ket{\Psi_A} =&e^{-\Gamma_St}\mathcal{E}_{+-}(t) + e^{-\Gamma_At}\mathcal{E}_{--}(t)-e^{- \gamma t}\mathcal{E}_0(t)        \label{eq:rhoAA_single_excitation_manifold}
\end{align}
with 
\begin{align}
    \mathcal{E}_0 = \frac{\kappa^4}{2}\sinh(2\Re[\Phi(t)]&),\\
    \mathcal{E}_{s_1s_2} =\frac14\bigg[ 1 + s_1\kappa^2+s_2&\big[\kappa^4\cosh(2\Re[\Phi(t)]) \nonumber\\&+ s_1\kappa^2\cos(2\Im[\Phi(t)])\big]\bigg].
\end{align}
    We make two key observations: First, due to the phonon interaction we observe transitions from the bare symmetric to the antisymmetric state, similar to the previous section. Moreover, we see an effect of the phonon bath long after its correlations have decayed: The dark and bright electronic states decay biexponentially with rates associated with the symmetric and antisymmetric polaronic Dicke states.

This biexponentiality can also be observed in the excitation number:
\begin{align}
    n(t) =& \bra{\Psi_S}{\rho}_S(t)\ket{\Psi_S} + \bra{\Psi_A}{\rho}_S(t)\ket{\Psi_A} \nonumber\\
    =&\frac{1}{2}\left(e^{-\Gamma_S t}(1+\kappa^2) +e^{-\Gamma_A t}(1-\kappa^2)\right)\label{eq:Number_PsiS}.
\end{align}
Thus, neither a perfectly bright nor a perfectly dark state exist in the presence of a vibrational coupling. This absence is twofold: First, one can see that the decay rate of the polaronic bright state is reduced compared to the pure atomic case, i.e.~$\Gamma_S <2\gamma$. 
Second, as a core result of this article, we make the following observation: 
The maximum decay rate $\Gamma_S$ as predicted in the polaron frame cannot be realized with ultrashort pulses. This is because an ultrashort pulse excites the bare electronic symmetric state, which inherits decay properties from the polaronic dark, as well as the polaronic bright state.

\subsubsection{Excitation with long pulses}
We now turn to excitation with a pulse duration that exceeds the phonon correlation time by far. Physically speaking, this means that the phonon environment equilibrates while the system is driven into the target state.
As our analytical expressions are not able to account for laser driving, we rely on our numerical method to obtain predictions for the mean photon number, as shown in Fig.~\ref{fig:PulseLengthDependenceNumber}. 

Comparing this with the results for short pulses one notices substantial differences: Instead of a significant mixing between bright and dark state after the initialization, as observed in Sec.~\ref{sec:2TLS_short_pulse}, we observe that this mixing can be heavily suppressed for long pulse durations by suppressing the dark state contribution. Indeed, we find that the excitation number following excitation can be approximately obtained as the solution of the equations of motion in the polaron frame with $\tilde{\rho}_S(0) = \ket{\Psi_S}\bra{\Psi_S}$, i.e., 
\begin{equation}
    n(t)= \exp(-\Gamma_S t),
\end{equation}
while differences can be attributed to the finite excitation time $\sigma$.

This behaviour can be interpreted in the following way: Since the dynamics induced by the laser is much slower than the dynamics of the phonon bath, the phonon environment undergoes an adiabatic displacement while the system is transferred into the excited state. This means that instead of the bare electronic symmetric state, the symmetric polaronic state gets excited by the laser. Thus, the system is governed by the polaron equations of motion, which allow for a mono-exponential decay.

\subsubsection{Initial state dependence}
As we have just demonstrated, for ultrashort pulses one obtains non-adiabatic excitation of bare electronic states, while one can adiabatically excite polaronic Dicke states with long pulses. This leads to a limitation in superradiant enhancement due to a mixing of polaronic bright and dark state occupations during initial polaron formation for short pulse durations. In this section we show that this limit persists if a different electronic state is initialized within the single excitation manifold. For this we investigate the dynamics of any possible single-excitation manifold state, given by ${\rho}_S(0) = {\rho}_{AA}(0)\ket{\Psi_A}\bra{\Psi_A} + {\rho}_{SS}(0)\ket{\Psi_S}\bra{{\Psi_S}}$, which can be prepared by controlling the phase between two lasers exciting two emitters placed in a planar waveguide in the perpendicular direction \cite{chu_subradiant_2022}. 

\begin{figure}
    \centering 
    \includegraphics[width=\linewidth]{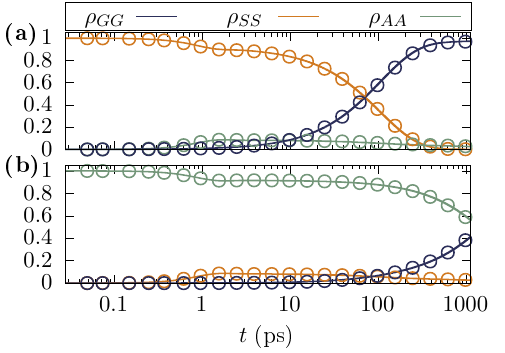}
    \caption{Occupations of the basis states $\ket{G_1,G_2}$, $\ket{\Psi_{S}}$, $\ket{\Psi_A}$ for (a) the initial state ${\rho}_S(0) = \ket{\Psi_S}\bra{\Psi_S}$, (b) the initial state ${\rho}_S(0) = \ket{\Psi_A}\bra{\Psi_A}$.}\label{fig:AnalyticStateOccupations}
\end{figure}

In Sec.~S3.4~of the SM \cite{supplement} we give the time evolution of an arbitrary initial state ${\rho}_S(0)$. We plot it for the special cases of ${\rho}_{SS}(0) = 1$ and ${\rho}_{AA}(0) = 1$ in figures \ref{fig:AnalyticStateOccupations}(a) and (b), respectively. Additionally, we show that the results can be reproduced with our numerically exact methods [cf. Sec.~S4~\cite{supplement}] by overlaying the analytical solutions with the outcome of our simulations.

We then obtain the mean excitation number in the system given for each initial state,
\begin{align}
    n(t) =& \rho_{SS}(0)\left(e^{\Gamma_S t}(1+\kappa^2) + e^{\Gamma_A t}(1-\kappa^2)\right)\nonumber\\
    &+\rho_{AA}(0)\left(e^{\Gamma_S t}(1-\kappa^2) + e^{\Gamma_A t}(1+\kappa^2)\right),\label{eq:Number_Arbitrary}
\end{align}
which generalizes Eq.~\eqref{eq:Number_PsiS}.

 While superradiant decay is not mono-exponential, we still want to estimate how long an excitation remains in the system. Thus, we define the lifetime $\tau$ as the time for which the mean excitation number has decayed to $1/e$, which equals the inverse decay rate for mono-exponential decay.
\begin{figure}
    \centering 
    \includegraphics[width=\linewidth]{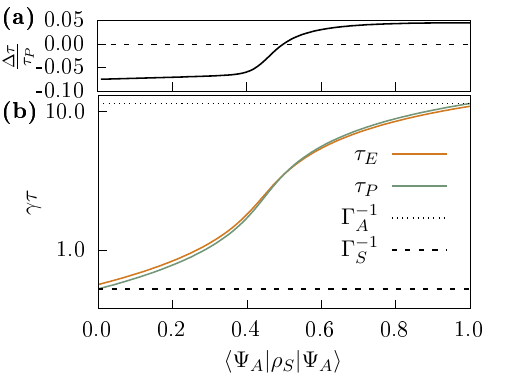}
    \caption{(a) Normalized difference $\Delta\tau = \tau_{P} - \tau_{E}$ between the predictions of polaronic excitations and bare electronic excitations for the initial state dependence of the lifetimes of single excitations. (b) Initial state dependence of excitation lifetimes $\tau_{E}$ (blue) and $\tau_{P}$ (orange) of an polaronic and electronic excitation, respectively. The lifetime values are normalized by the single emitter decay rate $\gamma$.}
    \caption{(a) Normalized difference $\Delta\tau = \tau_{P} - \tau_{E}$ between the predictions of polaronic excitations and bare electronic excitations for the initial state dependence of the lifetimes of single excitations. (b) Initial state dependence of excitation lifetimes $\tau_{E}$ (blue) and $\tau_{P}$ (orange) of an polaronic and electronic excitation, respectively. The lifetime values are normalized by the single emitter decay rate $\gamma$.}\label{fig:lifetime_limits}
\end{figure}
Fig.~\ref{fig:lifetime_limits}(b) shows the thus defined lifetimes $\tau_E$, if the system is prepared in the bare electronic state, and $\tau_P$, if prepared in a polaronic state, for different initial states. We observe differences, that become extremal for initialization in the bright and dark state, while the minimum (maximum) lifetime can still be achieved by initializing the system in the bright (dark) electronic state. For better visibility, we show the difference $\Delta\tau =\tau_P-\tau_E$ normalized to the polaronic lifetime in Fig.~\ref{fig:lifetime_limits}(a). 

Concluding, we find changes of several percent in the observed lifetime depending on whether the excitation is born in an electronic or polaronic state. Additionally, it is not possible to exceed the limiting rates dictated by the polaronic Dicke states regardless of the initialization.

\subsection{Pulse area dependence of the decay dynamics}\label{sec:2TLS_excitation_dependence}
We have found that for small pulse areas -- this means for negligible doubly excited state-populations -- the excitation pulse duration determines if the initialized system states are bare electronic, or polaronic states, or somewhere in between. This has an influence on the long-time dynamics of the two TLSs. The fact that the polaron transform does not change the doubly excited state occupation suggests that the situation will change if the doubly-excited state gets significantly occupied by increasing the pulse area of the laser pulse.

\begin{figure}
    \centering 
    \includegraphics[width=\linewidth]{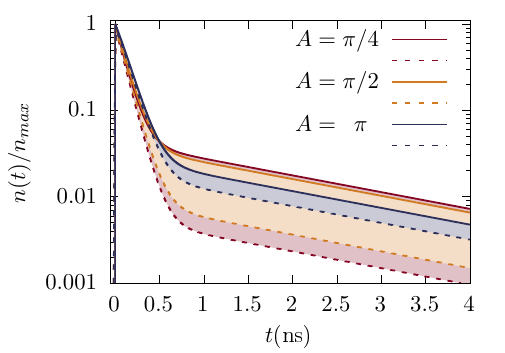}
    \caption{Normalized excitation number $n(t) / n_\text{max}$, where $n_\text{max} =\max_{t}n(t)$, for different laser pulses. We show excitations with different pulse areas $A$ for pulse lengths of $\tau=0.1$~ps (solid) and $\tau=20$~ps (dashed). The shaded areas indicate the resulting differences for different pulse lengths. }\label{fig:ExcitationDependenceBrightStatePulseArea}
\end{figure}

In Fig.~\ref{fig:ExcitationDependenceBrightStatePulseArea} we show the  excitation number dynamics of the two emitter system after pulsed excitation of the bright state for different pulse areas in two cases: We consider excitation with a pulse length of $\tau = 0.1$~ps, and excitation with $\tau = 20$~ps. To compare the cases of different pulse areas, we normalize the excitation number to its maximum right after the pulse. This is necessary as the maximum excitation number changes with the pulse area.

We observe that the difference between `bare electronic' and the `polaronic' initialization gets smaller the more the pulse area is increased. This can be understood by solving the polaronic and the bare electronic equations of motion for the initial state being the doubly excited state. We find that the time evolution for both the bare electronic as well as polaronic states is identical
\begin{subequations}
    \begin{align}
        \bra{X_1, X_2}{\rho}_{S}(t)&\ket{X_1, X_2}= e^{-(\Gamma_{XX})t},\\
        \bra{\Psi_S}{\rho}_{S}(t)&\ket{\Psi_S} =\frac{1}{2}\frac{\Gamma_S}{\Gamma_A}(1-e^{-\Gamma_A t})e^{-\Gamma_S t }(1+\kappa^2) \nonumber\\
        &+ \frac{1}{2}\frac{\Gamma_A}{\Gamma_S}(1-e^{-\Gamma_S t})e^{-\Gamma_A t }(1-\kappa^2),\\
        \bra{\Psi_A}{\rho}_{S}(t)&\ket{\Psi_A}=\frac{1}{2}\frac{\Gamma_S}{\Gamma_A}(1-e^{-\Gamma_A t})e^{-\Gamma_S t }(1-\kappa^2) \nonumber\\&+ \frac{1}{2}\frac{\Gamma_A}{\Gamma_S}(1-e^{-\Gamma_S t})e^{-\Gamma_A t }(1+\kappa^2),\\
        \bra{G_1, G_2}{\rho}_{S}(t)&\ket{G_1, G_2}= \frac{\Gamma_A^2 -\Gamma_A\Gamma_S+ \Gamma_S^2}{\Gamma_A\Gamma_S}e^{-(\Gamma_{XX} )t}\nonumber\\&-\frac{\Gamma_A}{\Gamma_S}e^{-\Gamma_A t}-\frac{\Gamma_S}{\Gamma_A}e^{-\Gamma_S t} + 1,
    \end{align}
\end{subequations}
with $\Gamma_{XX} = \Gamma_A + \Gamma_S = 2\gamma$. This is the immediate consequence of the doubly-excited state containing no inter-emitter coherences. Thus, it is frame independent and leads to identical dynamics in both frames. We obtain the mean photon number
\begin{align}
    n(t) =& 2e^{-(\Gamma_{XX})t} + \frac{\Gamma_A}{\Gamma_S}(1-e^{-\Gamma_S t})e^{-\Gamma_A t } \nonumber\\
    &+ \frac{\Gamma_S}{\Gamma_A}(1-e^{-\Gamma_A t})e^{-\Gamma_S t }.
\end{align}
Thus, we can explain the vanishing pulse length dependence for large pulse areas by a decreasing relevance of the initial single-excitation manifold occupations with increasing occupation of the doubly-excited state.
\section{Summary \& Discussion}\label{sec:summary}
\begin{figure}
    \centering
    \includegraphics{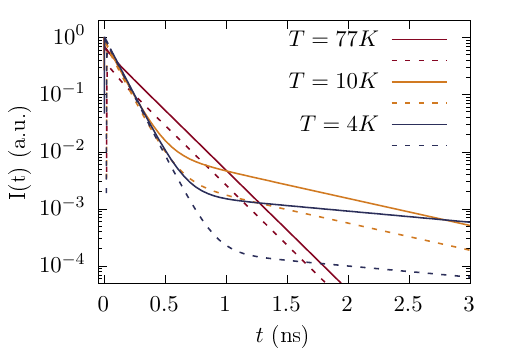}
    \caption{Normalized intensities of two superradiant quantum dots excited by a laser pulse at $t = 0$ with pulse area $A = \pi/8$ at different temperatures. The dashed lines represent driving with a pulse length of $\tau = 0.1$~ps and the solid lines represent $t = 20$~ps. For every temperature, normalization is performed with respect to the maximum intensity reached for $\tau = 0.1$~ps.}
    \label{fig:Intensity}
\end{figure}
Our goal for this article has been to investigate the influence of strongly coupled phonon environments on the superradiant decay of two dipoles.

Building up in complexity, we  first revisited a single emitter to show that its decay rate, within a flat spectral density assumption of the electromagnetic environment, is not impacted by the phonon environment. 
By contrast, inter-emitter coherences play a crucial role in superradiance, resulting in a complex interplay between the phonon and photon environments, even for a structureless electromagnetic environment. 
We have shown that the corresponding initial value problem can nonetheless be solved analytically up to the calculation of the environment correlation function, and we have presented simple expressions for the time evolution of two superradiant emitters.

To explore the intricate non-additive nature of three participating environments (one joint photon and two individual phonon ones) we numerically investigated the impact of excitation with a coherent resonant laser pulse. We have exposed a dependence of the superradiant lifetime on the pulse duration.
This can be attributed to the laser exciting the system adiabatically for long pulses, or non-adiabatically for shorter pulses: Adiabatic excitation leads to the excitation of the symmetric polaronic state which decays exponentially and minimizes the lifetime (in keeping with `fast' superradiant emission). By contrast, the bare electronic bright state, which is excited by short pulses, also possesses polaron frame dark state components that lead to an increase of its effective lifetime. 
We have shown that this behaviour is most pronounced in the low excitation regime where the frame-independent doubly-excited state is only negligibly occupied.

The analytic expressions presented in this article are valid as long as the spectral density does not lead to divergences under polaron transformation and for when the weak coupling approximation for the photon environment holds. Then, a super-Ohmic spectral density, which captures the dominant deformation potential coupling of excitons to phonons in self-assembled III-V quantum dots \cite{krummheuer_pure_2005, reiter_distinctive_2019}, and the vibrational environment of certain organic molecules \cite{clear_phonon-induced_2020}, leads to a  limited degree of mixing between polaronic dark and bright states (see Fig.~\ref{fig:AnalyticalDecoherence}). Different amounts of--or even complete--mixing of polaronic states can be expected for different choices of spectral density. More involved situations, e.g.~Ohmic spectral densities for which the polaron transform is not well defined, require other approaches for accurate predictions, such as a variational polaron transformation \cite{mccutcheon_general_2011}, or numerically exact methods \cite{gribben_exact_2022, cygorek_simulation_2022}.

In Fig.~\ref{fig:Intensity} we show the impact of more strongly coupled environments on the intensity, which is directly accessible in the experiment. To control the effective coupling strength we vary the temperature of the environment and compare short with longer excitation laser pulse duration. 
At low temperature, the bi-exponentiality becomes visible while it almost vanishes at high temperature. Instead, we observe a dynamical decoupling \cite{PhysRevLett.98.227403, nazir_modelling_2016} of the phonon environment for short excitation pulses, leading to an increased intensity compared to adiabatic excitation.

We expect that our qualitative conclusions regarding the influence of strong phonon coupling on superradiant emission are not limited to two emitters, but will persist in a similar fashion if the number of participating emitters is increased. However, such scaling up raises interesting broader questions about the robustness of cooperative emission to noise and disorder \cite{celardo_cooperative_2014, celardo_cooperative2_2014}, and about the validity of the weak coupling approximation for the electromagnetic environment.

Our results stress the physical relevance of the chosen frame in which approximate predictions are made. In our case, this manifests in the ability of the laser driving to switch between different frames, in which the resulting excited states feature different superradiant decay characteristics. Our article therefore introduces important concepts for further theoretical and experimental investigations of collective quantum effects in realistic condensed matter situations.
\bibliography{bibliography}
\clearpage


\section*{Supplementary material}

# Supplementary information: The role of polaron dressing in superradiant emission dynamics

J. Wiercinski[*], M. Cygorek, E. M. Gauger

July 12, 2024

## S1 Polaron frame master equation for superradiant decay

In this section we introduce the Hamilton operator for describing two emitters coupled to their individual phonon environments and a common electromagnetic environment. It can be decomposed into an undisturbed system Hamiltonian $H_S$, the phonon contribution, $H_{PN}$, and the photon contribution $H_{PT}$, i.e., $H = H_S + H_{PN} + H_{PT}$.

Both emitters are described as two-level systems (TLSs) at positions $\mathbf{R}_1$ and $\mathbf{R}_2$ with detunings $\omega_1$ and $\omega_2$ with respect to a reference frequency,

$$H_\text{S} = \hbar \sum_{j=1,2} \omega_j \sigma_j^+ \sigma_j^- \tag{S1}$$

with individual transition operators $\sigma_1^+ = |X_1\rangle\langle G_1|$, $\sigma_2^+ = |X_2\rangle\langle G_2|$, $\sigma_j^- = (\sigma_j^+)^\dagger$. We then describe the interaction of the two emitters with the photon environment within the rotating wave approximation. For the remainder of this derivation we assume that the emitters are very close together $|\mathbf{R}_1 - \mathbf{R}_2| \approx 0$, which leads to a coupling of the system to the photon environment via the collective operator $\sigma_S^+ = \frac{1}{\sqrt{2}}(\sigma_1^+ + \sigma_2^+)$, $\sigma_S^- = (\sigma_S^+)^\dagger$:

$$H_{PT} = \hbar \sum_\mathbf{k} \nu_\mathbf{k} a_\mathbf{k}^\dagger a_\mathbf{k} + \sqrt{2}\hbar \sum_\mathbf{k} g_\mathbf{k}(a_\mathbf{k}^\dagger \sigma_S^- + a_\mathbf{k} \sigma_S^+). \tag{S2}$$

The phonon coupling can be described using independent but identical boson baths [6, 7, 14]:

$$H_\text{PN} = \hbar \sum_{j,\mathbf{q}} \left[ \omega_\mathbf{q} b_{\mathbf{q},j}^\dagger b_{\mathbf{q},j} + \sigma_j^+ \sigma_j^- \xi_{\mathbf{q},j}(b_{\mathbf{q},j}^\dagger + b_{\mathbf{q},j}) \right] \tag{S3}$$

with the creation (annihilation) operator of the $\mathbf{q}$-mode of the $j$-th phonon environment $b_{\mathbf{q},j}^\dagger$ ($b_{\mathbf{q},j}$), the mode frequencies $\omega_\mathbf{q}$, and the phonon coupling strengths $\xi_{\mathbf{q},1} = \xi_{\mathbf{q},2}$. For

---

[*]jehw2000@hm.ac.uk

1

two phonon environments, the polaron transformation [5, 8] is performed individually for both environments by applying the operators $\mathcal{U}_j = e^{P_j}$, $j = 1, 2$, where

$$P_j = \sum_{\mathbf{q}} \sigma_+^j \sigma_-^j \left( \xi_{\mathbf{q},j} b_{\mathbf{q},j}^\dagger - \xi_{\mathbf{q},j}^* b_{\mathbf{q},j} \right)/\omega_{\mathbf{q}}, \tag{S4}$$

$$\mathcal{U}_j = \sigma_-^j \sigma_+^j + \sigma_+^j \sigma_-^j D_j, \tag{S5}$$

with the displacement operator for an individual emitter

$$D_j := \prod_{\mathbf{q}} \exp\left( (\xi_{\mathbf{q},j} b_{\mathbf{q},j}^\dagger - \xi_{\mathbf{q},j}^* b_{\mathbf{q},j})/\omega_{\mathbf{q}} \right). \tag{S6}$$

Applying the polaron transform to the Hamilton operator, $\tilde{H} = \mathcal{U}_1 \mathcal{U}_2 H \mathcal{U}_2^\dagger \mathcal{U}_1^\dagger$, we obtain

$$\tilde{H} = \hbar \sum_j \sigma_j^+ \sigma_j^- (\omega_j - \omega_j^R) + \hbar \sum_{\mathbf{k}} \nu_{\mathbf{k}} a_{\mathbf{k}}^\dagger a_{\mathbf{k}} + \hbar \sum_{j,\mathbf{q}} \omega_{\mathbf{q}} b_{\mathbf{q},j}^\dagger b_{\mathbf{q},j} + \hbar \sum_{j,\mathbf{k}} g_{\mathbf{k}} \left( a_{\mathbf{k}}^\dagger D_j^\dagger \sigma_j^- + a_{\mathbf{k}} D_j \sigma_j^+ \right). \tag{S7}$$

We now move into an interaction picture with respect to the uncoupled photon and phonon bath Hamiltonians and assume that the polaron-shifted emitters are identical and possess transition frequency zero, i.e. $\omega_1 - \omega_1^R = \omega_2 - \omega_2^R = 0$. The interaction picture Hamiltonian, $\tilde{H}_I$ then reads:

$$\tilde{H}_I = \hbar \sum_{j,\mathbf{k}} g_{\mathbf{k}} \left( a_{\mathbf{k}}^\dagger D_j^\dagger(t) \sigma_j^- e^{i\nu_{\mathbf{k}} t} + a_{\mathbf{k}} D_j(t) \sigma_j^+ e^{-i\nu_{\mathbf{k}} t} \right). \tag{S8}$$

We can now follow a similar master equation derivation for the reduced density operator $\tilde{\rho}_{S,PN} = \text{Tr}_{PT}[\tilde{\rho}]$, as presented in Sec. II of the main text for a single emitter. For this we evaluate the double commutator in

$$\frac{d\tilde{\rho}_{S,PN}}{dt}(t) = -\int_0^t ds \, \text{Tr}_{PT}\{[H_I(t), [H_I(t-s), \tilde{\rho}_{S,PN}(t-s)]]\} \tag{S9}$$

and make the flat density assumption $J_{PT}(\omega) = \sum_{\mathbf{k}} g_{\mathbf{k}}^2 \delta(\omega - \omega_{\mathbf{k}}) = \gamma/2\pi$. Consequently, we arrive at

$$\frac{d\tilde{\rho}_{S,PN}}{dt}(t) = \gamma \sum_{i,j=1,2} \left\{ D_i^\dagger(t) \sigma_i^- \tilde{\rho}_{S,PN}(t) D_j(t) \sigma_j^+ - \frac{1}{2} [D_i(t) \sigma_i^+ D_j^\dagger(t) \sigma_j^-, \tilde{\rho}_{S,PN}(t)]_+ \right\} \tag{S10}$$

with the anti-commutator $[A, B]_+$ of two arbitrary operators $A, B$.

Now, following standard polaron theory, we utilize that the residual phonon environment that is not captured by the polaron is weakly coupled to the polaronic degrees of freedom. This allows for describing the dynamics of the polaronic degrees of freedom by tracing over the phonon degrees of freedom in Eq. (S10) with the environment being in a Gibbs state [8, 10, 11, 13]. Consequently, we replace on the right-hand side of Eq. (S10).

$$D_i^\dagger(t) \tilde{\rho}_{S,PN}(t) D_j(t) \to \tilde{\rho}_{S,PN}(t) \text{Tr}[D_i^\dagger(t) \rho_{PN} D_j(t)] = \begin{cases} \tilde{\rho}_{S,PN}(t), & i = j \\ \kappa^2 \tilde{\rho}_{S,PN}(t), & i \neq j \end{cases} \tag{S11}$$

$$D_i(t) D_j^\dagger(t) \tilde{\rho}_{S,PN}(t) \to \text{Tr}[D_i(t) D_j^\dagger(t) \rho_{PN}] \tilde{\rho}_{S,PN}(t) \begin{cases} \tilde{\rho}_{S,PN}(t), & i = j \\ \kappa^2 \tilde{\rho}_{S,PN}(t), & i \neq j \end{cases} \tag{S12}$$

where we introduced the mean displacement $\kappa = \text{Tr}[D_{1/2}\rho_{PN}]$, with the Gibbs state density operator $\rho_{PN} = \frac{1}{\mathcal{Z}}e^{-\beta\hbar\sum_{j,\mathbf{q}}\omega_\mathbf{q} b^\dagger_{\mathbf{q},j} b_{\mathbf{q},j}}$, where $\mathcal{Z}$ is the partition sum [4]. Then, we obtain

$$\frac{d\tilde{\rho}_{S,PN}}{dt}(t) = \gamma \sum_{i=1,2} \left\{ \sigma_i^- \tilde{\rho}_{S,PN}(t)\sigma_i^+ - \frac{1}{2}\left[\sigma_i^+ \sigma_i^-, \tilde{\rho}_{S,PN}(t)\right]_+ \right\}$$
$$+ \gamma\kappa^2 \sum_{i \neq j} \left\{ \sigma_i^- \tilde{\rho}_{S,PN}(t)\sigma_j^+ - \frac{1}{2}\left[\sigma_i^+ \sigma_j^-, \tilde{\rho}_{S,PN}(t)\right]_+ \right\}. \tag{S13}$$

Now, one can introduce a Dicke basis in the polaron frame

$$|G_1, G_2\rangle := |G_1\rangle|G_2\rangle, \tag{S14a}$$

$$|\Psi_S\rangle := \frac{1}{\sqrt{2}}(|X_1\rangle|G_2\rangle + |G_1\rangle|X_2\rangle), \tag{S14b}$$

$$|\Psi_A\rangle := \frac{1}{\sqrt{2}}(|X_1\rangle|G_2\rangle - |G_1,\rangle|X_2\rangle), \tag{S14c}$$

$$|X_1, X_2\rangle := |X_1\rangle|X_2\rangle, \tag{S14d}$$

and introduce the transition operators $\sigma_{S/A}^+ = \frac{1}{\sqrt{2}}\left(\sigma_1^+ \pm \sigma_2^+\right)$. Equation (S13) then takes on Lindblad form [1, 4].

$$\frac{d\tilde{\rho}_{S,PN}}{dt}(t) = \Gamma_S \left( \sigma_S^- \tilde{\rho}_{S,PN}\sigma_S^+ - \frac{1}{2}\left[\sigma_S^+ \sigma_S^-, \tilde{\rho}_{S,PN}\right]_+ \right)$$
$$+ \Gamma_A \left( \sigma_A^- \tilde{\rho}_{S,PN}\sigma_A^+ - \frac{1}{2}\left[\sigma_A^+ \sigma_A^- \tilde{\rho}_{S,PN}\right]_+ \right) \tag{S15}$$

where we introduced the renormalized decay rates $\Gamma_{S/A} = \gamma(1 \pm \kappa^2)$.

## S2 Solutions of the polaron frame master equation for superradiant decay

In the previous section we derived the master equation (S15) for superradiant decay in the polaron frame for two emitters. Here, we sketch how to solve it. By projecting onto the basis states given by Eqs. (S14a)-(S14d) we obtain a set of sixteen equations of motion for the entries of the density operator, which describe polaronic Dicke states. These fall into two mutually independent groups. These are the equations of motion for the occupations

$$\frac{d}{dt}\langle X_1, X_2|\tilde{\rho}_{S,PN}|X_1, X_2\rangle = -(\Gamma_A + \Gamma_S)\langle X_1, X_2|\tilde{\rho}_{S,PN}|X_1, X_2\rangle, \tag{S16a}$$

$$\frac{d}{dt}\langle\Psi_S|\tilde{\rho}_{S,PN}|\Psi_S\rangle = \Gamma_S\langle X_1, X_2|\tilde{\rho}_{S,PN}|X_1, X_2\rangle - \Gamma_S\langle\Psi_S|\tilde{\rho}_{S,PN}|\Psi_S\rangle, \tag{S16b}$$

$$\frac{d}{dt}\langle\Psi_A|\tilde{\rho}_{S,PN}|\Psi_A\rangle = \Gamma_A\langle X_1, X_2|\tilde{\rho}_{S,PN}|X_1, X_2\rangle - \Gamma_A\langle\Psi_A|\tilde{\rho}_{S,PN}|\Psi_A\rangle, \tag{S16c}$$

$$\frac{d}{dt}\langle G_1, G_2|\tilde{\rho}_{S,PN}|G_1, G_2\rangle = \Gamma_A\langle\Psi_A|\tilde{\rho}_{S,PN}|\Psi_A\rangle + \Gamma_S\langle\Psi_S|\tilde{\rho}_{S,PN}|\Psi_S\rangle, \tag{S16d}$$

and the equations of motion for the off-diagonal elements

$$\frac{d}{dt}\langle\Psi_S|\tilde{\rho}_{S,PN}|\Psi_A\rangle = -\frac{\Gamma_S+\Gamma_A}{2}\langle\Psi_S|\tilde{\rho}_{S,PN}|\Psi_A\rangle, \tag{S17a}$$

$$\frac{d}{dt}\langle\Psi_S|\tilde{\rho}_{S,PN}|X_1,X_2\rangle = -\frac{2\Gamma_S+\Gamma_A}{2}\langle\Psi_S|\tilde{\rho}_{S,PN}|X_1,X_2\rangle, \tag{S17b}$$

$$\frac{d}{dt}\langle\Psi_A|\tilde{\rho}_{S,PN}|X_1,X_2\rangle = -\frac{2\Gamma_A+\Gamma_S}{2}\langle\Psi_A|\tilde{\rho}_{S,PN}|X_1,X_2\rangle, \tag{S17c}$$

$$\frac{d}{dt}\langle G_1,G_2|\tilde{\rho}_{S,PN}|X_1,X_2\rangle = -\frac{\Gamma_S+\Gamma_A}{2}\langle G_1,G_2|\tilde{\rho}_{S,PN}|X_1,X_2\rangle, \tag{S17d}$$

$$\frac{d}{dt}\langle G_1,G_2|\tilde{\rho}_{S,PN}|\Psi_S\rangle = \Gamma_S\langle\Psi_S|\tilde{\rho}_{S,PN}|X_1,X_2\rangle - \frac{\Gamma_S}{2}\langle G_1,G_2|\tilde{\rho}_{S,PN}|\Psi_S\rangle, \tag{S17e}$$

$$\frac{d}{dt}\langle G_1,G_2|\tilde{\rho}_{S,PN}|\Psi_A\rangle = \Gamma_A\langle\Psi_A|\tilde{\rho}_{S,PN}|X_1,X_2\rangle - \frac{\Gamma_A}{2}|G_1,G_2\rangle\tilde{\rho}_{S,PN}|\Psi_A\rangle, \tag{S17f}$$

as well as their complex conjugates.

The equation of motion for the doubly excited state occupation, Eq. (S16a), is solved by an exponential function

$$\langle X_1,X_2|\tilde{\rho}_{S,PN}(t)|X_1,X_2\rangle = \langle X_1,X_2|\tilde{\rho}_{S,PN}(0)|X_1,X_2\rangle e^{-(\Gamma_A+\Gamma_S)t}. \tag{S18a}$$

The Ansatz $\tilde{\rho}_{j,j}(t) = C(t)\exp(-\Gamma_j t)$, $j \in \{S,A\}$ solves Eqs. (S16b) and (S16c)

$$\begin{aligned}\langle\Psi_S|\tilde{\rho}_{S,PN}(t)|\Psi_S\rangle =& \langle\Psi_S|\tilde{\rho}_{S,PN}(0)|\Psi_S\rangle e^{-\Gamma_S t}\\ &+ \langle X_1,X_2|\tilde{\rho}_{S,PN}(0)|X_1,X_2\rangle\frac{\Gamma_S}{\Gamma_A}(1-e^{-\Gamma_A t})e^{-\Gamma_S t},\end{aligned} \tag{S18b}$$

$$\begin{aligned}\langle\Psi_A|\tilde{\rho}_{S,PN}(t)|\Psi_A\rangle =& \langle\Psi_A|\tilde{\rho}_{S,PN}(0)|\Psi_A\rangle e^{-\Gamma_A t}\\ &+ \langle X_1,X_2|\tilde{\rho}_{S,PN}(0)|X_1,X_2\rangle\frac{\Gamma_A}{\Gamma_S}(1-e^{-\Gamma_S t})e^{-\Gamma_A t}.\end{aligned} \tag{S18c}$$

The easiest way to obtain the ground state occupation is by using trace preservation.

$$\begin{aligned}&\langle G_1,G_2|\tilde{\rho}_{S,PN}(t)|G_1,G_2\rangle\\ &=1-\langle X_1,X_2|\tilde{\rho}_{S,PN}(t)|X_1,X_2\rangle-\langle\Psi_A|\tilde{\rho}_{S,PN}(t)|\Psi_A\rangle-\langle\Psi_S|\tilde{\rho}_{S,PN}(t)|\Psi_S\rangle\\ &= \tilde{\rho}_{X,X}(0)\left(1-e^{-(\Gamma_A+\Gamma_S)t}-\frac{\Gamma_A}{\Gamma_S}(1-e^{-\Gamma_S t})e^{-\Gamma_A t}-\frac{\Gamma_S}{\Gamma_A}(1-e^{-\Gamma_A t})e^{-\Gamma_S t}\right)\\ &\quad+\langle\Psi_A|\tilde{\rho}_{S,PN}(0)|\Psi_A\rangle\left(1-e^{-\Gamma_A t}\right)+\langle\Psi_S|\tilde{\rho}_{S,PN}(0)|\Psi_S\rangle\left(1-e^{-\Gamma_S t}\right)\\ &\quad+\langle G_1,G_2|\tilde{\rho}_{S,PN}(0)|G_1,G_2\rangle.\end{aligned} \tag{S18d}$$

Most equations of motions for the coherences are solved by exponential functions,

$$\langle\Psi_S|\tilde{\rho}_{S,PN}(t)|\Psi_A\rangle = \langle\Psi_S|\tilde{\rho}_{S,PN}(0)|\Psi_A\rangle e^{-\frac{\Gamma_S+\Gamma_A}{2}t}, \tag{S19a}$$

$$\langle\Psi_S|\tilde{\rho}_{S,PN}(t)|X_1,X_2\rangle = \langle\Psi_S|\tilde{\rho}_{S,PN}(0)|X_1,X_2\rangle e^{-\Gamma_S t}, \tag{S19b}$$

$$\langle\Psi_A|\tilde{\rho}_{S,PN}(t)|X_1,X_2\rangle = \langle\Psi_A|\tilde{\rho}_{S,PN}(0)|X_1,X_2\rangle e^{-\Gamma_A t}, \tag{S19c}$$

$$\langle G_1,G_2|\tilde{\rho}_{S,PN}(t)|X_1,X_2\rangle = \langle G_1,G_2|\tilde{\rho}_{S,PN}(0)|X_1,X_2\rangle e^{-\frac{\Gamma_S+\Gamma_A}{2}t}, \tag{S19d}$$

while two equations can be solved analogously to $\langle \Psi_S | \tilde{\rho}_{S,PN}(t) | \Psi_S \rangle$ and $\langle \Psi_S | \tilde{\rho}_{S,PN}(t) | \Psi_S \rangle$

$$\langle G_1, G_2 | \tilde{\rho}_{S,PN}(t) | \Psi_S \rangle = \langle G_1, G_2 | \tilde{\rho}_{S,PN}(0) | \Psi_S \rangle e^{-\frac{\Gamma_S}{2}t}$$
$$- \langle \Psi_S | \tilde{\rho}_{S,PN}(0) | X_1, X_2 \rangle (1 - e^{-\Gamma_S t}) e^{-\frac{\Gamma_S}{2}t}, \quad \text{(S19e)}$$

$$\langle G_1, G_2 | \tilde{\rho}_{S,PN}(t) | \Psi_A \rangle = \langle G_1, G_2 | \tilde{\rho}_{S,PN}(0) | \Psi_A \rangle e^{-\frac{\Gamma_A}{2}t}$$
$$- \langle \Psi_A | \tilde{\rho}_{S,PN}(0) | X_1, X_2 \rangle (1 - e^{-\Gamma_A t}) e^{-\frac{\Gamma_A}{2}t}. \quad \text{(S19f)}$$

The remaining contributions of the dynamical map can be obtained by complex conjugation.

Overall the set of equations Eqs. (S18a)-(S18d) and (S19a)-(S19f) can be perceived as a linear map in Liouville space

$$\tilde{\rho}_{S,PN}(t) = \mathcal{M}(t) \tilde{\rho}_{S,PN}(0). \quad \text{(S20)}$$

## S3 Superradiant decay with bare electronic initial states

In this section, we show how to calculate the time evolution of the occupations of the bare electronic Dicke states during superradiant decay. For this, we assume that the initial state of the excitonic system $\rho_S(0)$ and its phonon environment are separable, i.e. $\rho_{S,PN}(0) = \rho_S(0) \otimes \rho_{PN}(0)$, and the phonon state $\rho_{PN}(0)$ is a Gibbs state. As we show in the main text, these capture superradiant decay characteristics for system initialized using a laser pulse with a pulse length much shorter than the correlation time of the environment.

This section is organised as follows: First, we sketch how one obtains the time evolution of the bare electronic states from the dynamics of the polaronic Dicke states. Then, in section S3.2, we derive a general form of the relevant correlation functions of the combined phonon environments and derive expressions which only depend on correlations a a single phonon environment. We then calculate these expressions in S3.3 up to the point of evaluating the environment correlation function. We summarize and condense our results in Sec. S3.4. Finally, we solve Eq. (S22) for different initial conditions, first, for the trivial propagator $\mathcal{M}(t) = \mathbb{1}$ (Sec. S3.5), which corresponds to free decay of coherences, and then for superradiant decay (Sec. S3.6).

### S3.1 Bare state dynamics from the polaron frame equations of motion

The general idea for obtaining the reduced density operator for bare electronic states at time $t$ is to take advantage of the dynamical map of the combined system and phonon degrees of freedom in the polaron frame, $\mathcal{M}(t)$, which we calculated in Sec. S2. Then, the density operator for the bare electronic degrees of freedom can be written as

$$\rho_S(t) = \text{Tr}_{PN} \left[ \mathcal{T}^{-1}(t) \left\{ \mathcal{M}(t) \mathcal{T}[\rho_S(0) \otimes \rho_{PN}(0)] \right\} \right], \quad \text{(S21)}$$

where $\mathcal{T}[\rho] = \mathcal{U}_1 \mathcal{U}_2 \rho \mathcal{U}_2^\dagger \mathcal{U}_1^\dagger$ and $\mathcal{T}^{-1}[\tilde{\rho}] = \mathcal{U}_1^\dagger \mathcal{U}_2^\dagger \tilde{\rho} \mathcal{U}_2 \mathcal{U}_1$ are the polaron transform and inverse polaron transform represented in Liouville space. Eq. (S21) corresponds to transforming

the initial density operator in the interaction picture into the polaron frame, applying the dynamical map to propagate the system to time $t$, and then performing the inverse polaron transform. The time-evolved back transform takes into account the time evolution of the environment degrees of freedom while the system evolves in the polaron frame due to the action of the propagator

$$\mathcal{T}^{-1}(t) = e^{\frac{itH_E}{\hbar}} \mathcal{T}^{-1}[\tilde{\rho}] e^{-\frac{itH_E}{\hbar}} = \mathcal{U}_1^\dagger(t) \mathcal{U}_2^\dagger(t) \tilde{\rho} \mathcal{U}_2(t) \mathcal{U}_1(t), \tag{S22}$$

where $\mathcal{U}_{1/2}(t) = e^{\frac{itH_E}{\hbar}} \mathcal{U}_{1/2} e^{-\frac{itH_E}{\hbar}}$. Here, the Hamilton operator $H_E = \hbar \sum_{j=1,2} \sum_{\mathbf{q}} \omega_{\mathbf{q}} b_{\mathbf{q},j}^\dagger b_{\mathbf{q},j}$ describes the free evolution of the phonon degrees of freedom. Only at the end, the trace over the phonon degrees of freedom is taken. This leaves phonon correlations intact throughout the time evolution.

We represent the combined action of the polaron transform and the back transform, in the Dicke basis of delocalized two-emitter states

$$\begin{aligned}
\mathcal{U}_1 \mathcal{U}_2 =& |G_1, G_2\rangle\langle G_1, G_2| + (|\Psi_S\rangle\langle\Psi_S| + |\Psi_A\rangle\langle\Psi_A|)\mathcal{D}_+ + (|\Psi_S\rangle\langle\Psi_A| + |\Psi_A\rangle\langle\Psi_S|)\mathcal{D}_- \\
& + |X_1, X_2\rangle\langle X_1, X_2| D_1 D_2,
\end{aligned} \tag{S23a}$$

$$\begin{aligned}
\mathcal{U}_2^\dagger \mathcal{U}_1^\dagger =& |G_1, G_2\rangle\langle G_1, G_2| + (|\Psi_S\rangle\langle\Psi_S| + |\Psi_A\rangle\langle\Psi_A|)\mathcal{D}_+^\dagger + (|\Psi_S\rangle\langle\Psi_A| + |\Psi_A\rangle\langle\Psi_S|)\mathcal{D}_-^\dagger \\
& + |X_1, X_2\rangle\langle X_1, X_2| D_1^\dagger D_2^\dagger,
\end{aligned} \tag{S23b}$$

with $\mathcal{D}_\pm(t) = \frac{1}{2}(D_1 \pm D_2)$.

### S3.2 Properties of the correlation functions for two emitters

In this section we present how the action of the (inverse) polaron transformation at zero time (time $t$) can be captured via two-time correlations of displacement operators of individual phonon baths. Equations (S23a) and (S23b) in combination with Eq. (S21) imply that the phonon influence on the density operator is determined by linear combinations of expectation values of various combinations of four of the phonon environment operators $\langle \mathcal{O}_1^\dagger(0) \mathcal{O}_2(t) \mathcal{O}_3^\dagger(t) \mathcal{O}_4(0) \rangle_{\rho_{PN}}$, with $\mathcal{O}_j \in \{\mathbb{1}, \mathcal{D}_\pm, D_1 D_2\}$ for $j \in \{1, 2, 3, 4\}$, where $\mathcal{O}(t) := \exp(itH_E/\hbar) \mathcal{O} \exp(-itH_E/\hbar)$.

For all cases presented in the main text, only a small subset, i.e. the expectation values $\langle \mathcal{D}_{s_1}^\dagger(0) \mathcal{D}_{s_2}(t) \mathcal{D}_{s_3}^\dagger(t) \mathcal{D}_{s_4}(0) \rangle$ with $s_1, s_2, s_3, s_4 \in \{+, -\}$, as well as $\langle \mathcal{D}_{s_1}^\dagger(0) \mathcal{D}_{s_2}(0) \rangle$, are relevant, and here, we only derive analytical expressions for these. Other forms of correlations, e.g., $\langle \mathcal{D}_+^\dagger(0) \mathcal{D}_+(t) \mathcal{D}_-^\dagger(t) \rangle$, may appear for other time evolutions $\mathcal{M}(t)$. This section is aimed at giving a clear understanding of how they can easily be evaluated. For this we show that, as long as the environments are independent and identical, all environment correlations can be written in terms of the correlation functions

$$C_+(t) := \langle D_j(t) \rangle = (\langle D_j^\dagger(t) \rangle)^*, \tag{S24a}$$

$$C_{-+}(t) = \langle D_j^\dagger(t) D_j(0) \rangle = (\langle D_j^\dagger(0) D_j(t) \rangle)^*, \tag{S24b}$$

$$C_{++}(t) = \langle D_j(t) D_j(0) \rangle = (\langle D_j^\dagger(0) D_j^\dagger(t) \rangle)^*, \tag{S24c}$$

$$C_{-++}(t) = \langle D_j^\dagger(0) D_j(t) D_j(0) \rangle = (\langle D_j^\dagger(0) D_j^\dagger(t) D_j(0) \rangle)^* \tag{S24d}$$

of individual environments, which are labelled by $j \in \{1, 2\}$. For an explanation on how to calculate these correlations, we refer to the next section S3.3 of this supplement.

We start this section by first introducing basic conversion rules for the phonon environment operators and then, building our way up to more complex expressions. To

shorten notation, we omit the time argument of the operators, if they are evaluated at $t = 0$.

We start with the product of the environment operator combinations and their Hermitean conjugates,

$$\mathcal{D}_\pm(t)\mathcal{D}_\pm^\dagger(t) = \frac{1}{4}\left(2 \pm (D_1(t)D_2^\dagger(t) + D_2(t)D_1^\dagger(t))\right), \tag{S25a}$$

which shows that these operators are not Hermitean. Similarly, one obtains, by straightforwardly inserting the definition of $\mathcal{D}_\pm(t)$,

$$\mathcal{D}_+(t)\mathcal{D}_+^\dagger(t) + \mathcal{D}_-(t)\mathcal{D}_-^\dagger(t) = 1 \tag{S25b}$$

$$\mathcal{D}_\pm(t)\mathcal{D}_\mp^\dagger(t) = \pm\frac{1}{4}(D_2(t)D_1^\dagger(t) - D_1(t)D_2^\dagger(t)) \tag{S25c}$$

$$\mathcal{D}_+^\dagger(t)\mathcal{O}\mathcal{D}_+(t) = \mathcal{D}_-^\dagger(t)\mathcal{O}\mathcal{D}_-(t) + \frac{1}{2}(D_1^\dagger(t)\mathcal{O}D_2(t) + D_2^\dagger(t)\mathcal{O}D_1(t)) \tag{S25d}$$

$$\mathcal{D}_+^\dagger(t)\mathcal{O}\mathcal{D}_-(t) = \mathcal{D}_-^\dagger(t)\mathcal{O}\mathcal{D}_+(t) + \frac{1}{2}(D_2^\dagger(t)\mathcal{O}D_1(t) - D_1^\dagger(t)\mathcal{O}D_2(t)), \tag{S25e}$$

where $\mathcal{O}$ denotes an arbitrary phonon environment operator. From Eq. (S25c) we directly obtain $\mathcal{D}_{s_1}^\dagger \mathcal{D}_{s_2}(t)\mathcal{D}_{s_3}^\dagger(t)\mathcal{D}_{s_4} = -\mathcal{D}_{s_1}^\dagger \mathcal{D}_{s_3}(t)\mathcal{D}_{s_2}^\dagger(t)\mathcal{D}_{s_4}$, for $s_2 \neq s_3$. We now use these rules to break down every combination of environment operators into correlation functions containing only $D_1(t)$ and $D_2(t)$.

We begin with correlations of the form $\langle \mathcal{D}_{s_1}^\dagger \mathcal{D}_{s_2}(t)\mathcal{D}_{s_2}^\dagger(t)\mathcal{D}_{s_1}\rangle$, $s_1, s_2 \in \{+, -\}$. For the fully symmetric contribution we insert the definition of $\mathcal{D}_+(t)$ and utilize (S25a):

$$\langle \mathcal{D}_+^\dagger \mathcal{D}_+(t)\mathcal{D}_+^\dagger(t)\mathcal{D}_+\rangle = \frac{1}{4}\langle \mathcal{D}_+^\dagger(2 + D_1(t)D_2^\dagger(t) + D_2(t)D_1^\dagger(t))\mathcal{D}_+\rangle$$
$$= \frac{1}{4}(1 + |C_+(t)|^2) + \frac{1}{4}\Re[C_{-++}(t)C_-(t)] + \frac{1}{8}\left(|C_{-+}(t)|^2 + |C_{++}(t)|^2\right). \tag{S26}$$

Combining this result with Eq. (S25b), one obtains

$$\langle \mathcal{D}_+^\dagger \mathcal{D}_-(t)\mathcal{D}_-^\dagger(t)\mathcal{D}_+\rangle = \langle \mathcal{D}_+^\dagger \mathcal{D}_+\rangle - \langle \mathcal{D}_+^\dagger \mathcal{D}_+(t)\mathcal{D}_+^\dagger(t)\mathcal{D}_+\rangle$$
$$= \frac{1}{4}(1 + |C_+(t)|^2) - \frac{1}{4}\Re[C_{-++}(t)C_-(t)] + \frac{1}{8}(|C_{-+}(t)|^2 + |C_{++}(t)|^2), \tag{S27}$$

and from this, with property Eq. (S25d),

$$\langle \mathcal{D}_-^\dagger \mathcal{D}_-(t)\mathcal{D}_-^\dagger(t)\mathcal{D}_-\rangle = \langle \mathcal{D}_+^\dagger \mathcal{D}_-(t)\mathcal{D}_-^\dagger(t)\mathcal{D}_+\rangle$$
$$\quad - \frac{1}{2}\left(\langle D_1^\dagger \mathcal{D}_-(t)\mathcal{D}_-^\dagger(t)D_2\rangle + \langle D_2^\dagger \mathcal{D}_-(t)\mathcal{D}_-^\dagger(t)D_1\rangle\right)$$
$$= \frac{1}{4}(1 - |C_+(t)|^2) - \frac{1}{4}\Re[C_{-++}(t)C_-(t)] - \frac{1}{8}(|C_{-+}(t)|^2 + |C_{++}(t)|^2). \tag{S28}$$

And, finally, we get

$$\langle \mathcal{D}_-^\dagger \mathcal{D}_+(t)\mathcal{D}_+^\dagger(t)\mathcal{D}_-\rangle = \frac{1}{2}\langle \mathcal{D}_-^\dagger \mathcal{D}_-\rangle + \frac{1}{2}\Re[\langle \mathcal{D}_-^\dagger D_1(t)D_2^\dagger(t)\mathcal{D}_-\rangle]$$
$$= \frac{1}{2}(1 - |C_+(t)|^2) + \frac{1}{4}\Re[C_{-++}(t)C_-(t)] - \frac{1}{8}(|C_{++}(t)|^2 + |C_{-+}(t)|^2). \tag{S29}$$

7

Now, we investigate the cases $\langle \mathcal{D}^\dagger_{s_1} \mathcal{D}^\dagger_{s_2}(t) \mathcal{D}^\dagger_{-s_2}(t) \mathcal{D}_{-s_1} \rangle$, $s_1, s_2 \in \{+, -\}$. We begin with

$$\langle \mathcal{D}^\dagger_- \mathcal{D}_\pm(t) \mathcal{D}^\dagger_\mp(t) \mathcal{D}_+ \rangle = \pm \frac{1}{4} \langle \mathcal{D}^\dagger_- (D_2(t) D_1^\dagger(t) - D_1(t) D_2^\dagger(t)) \mathcal{D}_+ \rangle$$
$$= \pm \frac{1}{8} (2i\Im[C^*_{-++}(t) C_+(t)] + |C_{++}(t)|^2 - |C_{-+}(t)|^2), \quad \text{(S30)}$$

where we replaced $\mathcal{D}^\dagger_-$ and $\mathcal{D}_+$ by their definitions and calculate the individual terms using Eqs. (S24a)-(S24d). We can use this result, together with Eq. (S25c) and Eq. (S25e) to determine

$$\langle \mathcal{D}^\dagger_+ \mathcal{D}_\pm(t) \mathcal{D}^\dagger_\mp(t) \mathcal{D}_- \rangle = \pm \frac{1}{8} (2i\Im[C^*_{-++}(t) C_+(t)] - |C_{++}(t)|^2 + |C_{-+}(t)|^2). \quad \text{(S31)}$$

### S3.3 Calculation of the single-environment correlation functions

In the previous section we have shown that the influence of the two photon environments can be described by correlation functions of multiple time-dependent displacements of individual emitters, $D_1(t)$ and $D_2(t)$. In this section we relate them to one single quantity, the environment correlation function

We make use of relations for a general displacement operator of a single boson mode, $\mathcal{B}_{\mathbf{q},j}(\alpha) = \exp(\alpha b^\dagger_{\mathbf{q},j} - \alpha^* b_{\mathbf{q},j})$, where $\alpha \in \mathbb{C}$ is an arbitrary complex number. To this end, we write the polaron displacement operator as

$$D_j(0) = \sum_{\mathbf{q}} \mathcal{B}_{\mathbf{q},j}(\xi_{\mathbf{q}}/\omega_{\mathbf{q}}). \quad \text{(S32a)}$$

The consecutive application of displacement operators can be expressed as a single displacement

$$\mathcal{B}_{\mathbf{q},j}(\alpha) \mathcal{B}_{\mathbf{q},j}(\beta) = \mathcal{B}_{\mathbf{q},j}(\alpha + \beta) e^{i\Im(\alpha\beta^*)} \quad \text{(S32b)}$$

$$\mathcal{B}^\dagger_{\mathbf{q},j}(\alpha) = \mathcal{B}_{\mathbf{q},j}(-\alpha). \quad \text{(S32c)}$$

When evolving in time, the displaced environment acquires a phase, i.e.

$$e^{i\omega_{\mathbf{q}} t b^\dagger_{\mathbf{q},j} b_{\mathbf{q},j}} \mathcal{B}_{\mathbf{q},j}(\alpha) e^{-i\omega_{\mathbf{q}} t b^\dagger_{\mathbf{q},j} b_{\mathbf{q},j}} = \mathcal{B}_{\mathbf{q},j}(\alpha e^{i\omega_{\mathbf{q}} t}). \quad \text{(S32d)}$$

Additionally, the expectation value of the displacement operator for a Gibbs state is known to be [8]

$$\text{Tr}\left[\mathcal{B}_{\mathbf{q},j}(\alpha) \exp(-\beta \omega_{\mathbf{q}} b^\dagger_{\mathbf{q},j} b_{\mathbf{q},j})\right] = \exp\left(-\frac{|\alpha|^2}{2} \coth(\beta\omega_{\mathbf{q}}/2)\right). \quad \text{(S32e)}$$

Properties Eqs. (S32a)-(S32e) are sufficient to calculate all correlations appearing in the time evolution of the two-emitter problem, as presented in Sec. S3.2. In the remainder of the section we demonstrate the process of calculating these expectation values using examples of one, two, and three-operator expectation values and give expressions for the remaining ones.

We start be calculating the expectation value of the time dependent polaron displacement operator, the mean displacement.

$$C_+(t) = \langle D_j(t) \rangle = \prod_{\mathbf{q}} \left\langle \mathcal{B}_{\mathbf{q},j}\left(\frac{\xi_{\mathbf{q}}}{\omega_{\mathbf{q}}} e^{i\omega_{\mathbf{q}} t}\right) \right\rangle = \prod_{\mathbf{q}} \exp\left(-\frac{1}{2} \frac{\xi^2_{\mathbf{q}}}{\omega^2_{\mathbf{q}}} \left|e^{i\omega_{\mathbf{q}} t}\right|^2 \coth(\beta\omega_{\mathbf{q}}/2)\right)$$
$$= \exp\left(-\frac{1}{2} \int d\omega \frac{J_{PN}(\omega)}{\omega^2} \coth(\beta\omega/2)\right). \quad \text{(S33)}$$

Being time-independent, we refer to it as $\kappa := C_+(t)$ in the main text. Here, we introduced the phonon spectral density

$$J_{PN}(\omega) = \sum_{\mathbf{q}} \xi_{\mathbf{q}}^2 \delta(\omega - \omega_{\mathbf{q}}). \tag{S34}$$

We now turn to the two-times correlation function $C_{-+}(t) = \langle D_j^\dagger(t) D_j(0) \rangle$.

$$C_{-+}(t) = \prod_{\mathbf{q}} \left\langle \mathcal{B}_{\mathbf{q},j}^\dagger \left( \frac{\xi_q}{\omega_q} e^{i\omega_{\mathbf{q}} t} \right) \mathcal{B}_{\mathbf{q},j} \left( \frac{\xi_{\mathbf{q}}}{\omega_{\mathbf{q}}} \right) \right\rangle$$
$$= \kappa^2 \exp\left( \int d\omega \frac{J(\omega)}{\omega^2} \left[ \cos(\omega t) \coth(\beta\omega/2) - i \sin(\omega t) \right] \right) = \kappa^2 \exp(\Phi(t)), \tag{S35}$$

where we defined the environment correlation function

$$\Phi(t) = \int d\omega \frac{J(\omega)}{\omega^2} \left[ \cos(\omega t) \coth(\beta\omega/2) - i \sin(\omega t) \right]. \tag{S36}$$

Analogously, one finds for Eq. (S24c)

$$C_{++}(t) = \langle D_j(t) D_j(0) \rangle = \kappa^2 \exp(-\Phi(t)). \tag{S37}$$

Three operator correlations given by Eq. (S24d) can be readily obtained, as well.

$$C_{-++}(t) = \langle D_j^\dagger(0) D_j(t) D_j(0) \rangle = \kappa \prod_{\mathbf{q}} \exp\left( 2i \frac{\xi_{\mathbf{q}}^2}{\omega_{\mathbf{q}}^2} \sin(\omega_{\mathbf{q}} t) \right) = \kappa \exp(-2i\Im[\Phi(t)]). \tag{S38}$$

### S3.4 Summary

With the results of the previous two Sections we find a more concise expression for the environment correlations:

$$\langle \mathcal{D}_+^\dagger \mathcal{D}_+(t) \mathcal{D}_+^\dagger(t) \mathcal{D}_+ \rangle = \frac{1}{4}\left(1 + \kappa^2\right) + \frac{1}{4}\left(\kappa^4 \cosh(2\Re[\Phi(t)]) + \kappa^2 \cos(2\Im[\Phi(t)])\right), \tag{S39a}$$

$$\langle \mathcal{D}_+^\dagger \mathcal{D}_-(t) \mathcal{D}_-^\dagger(t) \mathcal{D}_+ \rangle = \frac{1}{4}\left(1 + \kappa^2\right) - \frac{1}{4}\left(\kappa^4 \cosh(2\Re[\Phi(t)]) + \kappa^2 \cos(2\Im[\Phi(t)])\right), \tag{S39b}$$

$$\langle \mathcal{D}_-^\dagger \mathcal{D}_-(t) \mathcal{D}_-^\dagger(t) \mathcal{D}_- \rangle = \frac{1}{4}\left(1 - \kappa^2\right) + \frac{1}{4}\left(\kappa^4 \cosh(2\Re[\Phi(t)]) - \kappa^2 \cos(2\Im[\Phi(t)])\right), \tag{S39c}$$

$$\langle \mathcal{D}_-^\dagger \mathcal{D}_+(t) \mathcal{D}_+^\dagger(t) \mathcal{D}_- \rangle = \frac{1}{4}\left(1 - \kappa^2\right) - \frac{1}{4}\left(\kappa^4 \cosh(2\Re[\Phi(t)]) - \kappa^2 \cos(2\Im[\Phi(t)])\right), \tag{S39d}$$

$$\langle \mathcal{D}_+^\dagger \mathcal{D}_\pm(t) \mathcal{D}_\mp^\dagger(t) \mathcal{D}_- \rangle = \pm\frac{1}{4}\left(\kappa^2 i \sin(2\Im[\Phi(t)]) + \kappa^4 \sinh(2\Re[\Phi(t)])\right), \tag{S39e}$$

$$\langle \mathcal{D}_-^\dagger \mathcal{D}_\pm(t) \mathcal{D}_\mp^\dagger(t) \mathcal{D}_+ \rangle = \pm\frac{1}{4}\left(\kappa^2 i \sin(2\Im[\Phi(t)]) - \kappa^4 \sinh(2\Re[\Phi(t)])\right), \tag{S39f}$$

which can be summarized for $s_1, s_2 \in \{+,-\}$ and $s_4 \neq s_1, s_3 \neq s_2$ as

$$\langle \mathcal{D}_{s_1}^\dagger \mathcal{D}_{s_2}(t) \mathcal{D}_{s_2}^\dagger(t) \mathcal{D}_{s_1} \rangle = \frac{1}{4}\left(1 + s_1 \kappa^2\right) + \frac{s_1 s_2}{4}\left(\kappa^4 \cosh(2\Re[\Phi(t)]) + s_1 \kappa^2 \cos(2\Im[\Phi(t)])\right) \tag{S40}$$

$$\langle \mathcal{D}_{s_1}^\dagger \mathcal{D}_{s_2}(t) \mathcal{D}_{s_3}^\dagger(t) \mathcal{D}_{s_4} \rangle = \frac{1}{4}\left(s_2 \kappa^2 i \sin(2\Im[\Phi(t)]) + s_1 s_2 \kappa^4 \sinh(2\Re[\Phi(t)])\right). \tag{S41}$$

9

## S3.5  Analytical solution for decoherence without decay

In this section, we derive the phonon influence in absence of radiative decay by choosing $\mathcal{M}(t) = \mathbb{1}$. This corresponds to two emitters being initialized in a bare electronic state and then evolving only due to their phonon coupling. Equation (S21) can then be written as

$$\rho_S(t) = \text{Tr}_{PN}[\mathcal{U}_1^\dagger(t)\mathcal{U}_2^\dagger(t)\mathcal{U}_1\mathcal{U}_2\rho_S(0)\rho_{PN}(0)\mathcal{U}_1^\dagger\mathcal{U}_2^\dagger\mathcal{U}_1(t)\mathcal{U}_2(t)]. \tag{S42}$$

We evaluate this expression for an arbitrary diagonal density operator in the Dicke basis. It is easy to see that for $\rho_S(0) = |G_1, G_2\rangle\langle G_1, G_2|$

$$\rho_S(t) = |G_1, G_2\rangle\langle G_1, G_2|, \tag{S43}$$

and

$$\rho_S(t) = |X_1, X_2\rangle\langle X_1, X_2|\langle D_1^\dagger D_2^\dagger D_1(t)D_2(t)D_1^\dagger(t)D_2^\dagger(t)D_1 D_2\rangle = |X_1, X_2\rangle\langle X_1, X_2|, \tag{S44}$$

for $\rho_S(0) = |X_1, X_2\rangle\langle X_1, X_2|$, as the phonon coupling does not lead to transition between manifolds with different electronic excitation.

Within the single-excitation manifold, however, the polaron transform leads to a mixing between the states $|\Psi_S\rangle$ and $|\Psi_A\rangle$, resulting in transitions from the initialised to the initially unoccupied state. One obtains for $\rho_S(0) = \rho_{SS}(0)|\Psi_S\rangle\langle\Psi_S| + \rho_{AA}(0)|\Psi_A\rangle\langle\Psi_A|$

$$\rho_S(t) = \left(\sum_{s_1,s_2\in\{\pm\}} \langle \mathcal{D}_{s_1}^\dagger \mathcal{D}_{s_1}(t)\mathcal{D}_{s_2}^\dagger(t)\mathcal{D}_{s_2}\rangle\right) (\rho_{SS}(0)|\Psi_S\rangle\langle\Psi_S| + \rho_{AA}(0)|\Psi_A\rangle\langle\Psi_A|)$$

$$+ \left(\sum_{s_1,s_2\in\{\pm\}} \langle \mathcal{D}_{s_1}^\dagger \mathcal{D}_{-s_1}(t)\mathcal{D}_{s_2}^\dagger(t)\mathcal{D}_{-s_2}\rangle\right) (\rho_{SS}(0)|\Psi_A\rangle\langle\Psi_A| + \rho_{AA}(0)|\Psi_S\rangle\langle\Psi_S|). \tag{S45}$$

We can now insert the expressions for the environment correlations as derived in Sec. S3.2:

$$\left(\sum_{s_1,s_2\in\{\pm\}}\langle \mathcal{D}_{s_1}^\dagger \mathcal{D}_{s_1}(t)\mathcal{D}_{s_2}^\dagger(t)\mathcal{D}_{s_2}\rangle\right) = \frac{1}{2} + \frac{\kappa^4}{2}\cosh(2\Re[\Phi(t)]) + \frac{\kappa^4}{2}\sinh(2\Re[\Phi(t)])$$

$$= \frac{1}{2} + \frac{\kappa^4}{2}\exp(2\Re[\Phi(t)]) =: a_+(t), \tag{S46a}$$

$$\left(\sum_{s_1,s_2\in\{\pm\}}\langle \mathcal{D}_{s_1}^\dagger \mathcal{D}_{-s_1}(t)\mathcal{D}_{s_2}^\dagger(t)\mathcal{D}_{-s_2}\rangle\right) = \frac{1}{2} - \frac{\kappa^4}{2}\cosh(2\Re[\Phi(t)]) - \frac{\kappa^4}{2}\sinh(2\Re[\Phi(t)])$$

$$= \frac{1}{2} - \frac{\kappa^4}{2}\exp(2\Re[\Phi(t)]) =: a_-(t). \tag{S46b}$$

## S3.6  Analytical solutions for superradiant decay

In this section we present the derivation of the state occupations during superradiant decay for different initial states of the system. For this we again need to solve Eq. (S21). In contrast to the previous section, this time the linear map $\mathcal{M}(t)$ in the polaron frame is not trivial, but instead determined by the equations of motions in the

polaron frame as has been derived in Sec. S2. We refer to the bare electronic state occupations as $\langle X_1, X_2|\rho_S|X_1, X_2\rangle := \rho_{XX}$, $\langle \Psi_S|\rho_S|\Psi_S\rangle := \rho_{SS}$, $\langle \Psi_A|\rho_S|\Psi_A\rangle = \rho_{AA}$, $\langle G_1, G_2|\rho_S|G_1, G_2\rangle = \rho_{GG}$. Correspondingly, for $i,j,k,l \in \{G, S, A, X\}$, we write the linear map component that maps the state $\rho_{kl}(0)$ to $\rho_{ij}(t)$ as $M_{ijkl}(t)$.

After having found the dynamical map in the polaron frame as a matrix operator in the previous section, we can proceed by calculating the lab frame time evolution as given by Eq. (S21). For the doubly excited state being the initial state, i.e. $\rho_S(0) = |X_1, X_2\rangle\langle X_1, X_2|$, we find

$$\rho_{XX}(t) = M_{XXXX}(t)\left(\sum_{s_1 \in \{\pm\}}\langle \mathcal{D}^\dagger_{s_1}\mathcal{D}_{s_1}(t)\mathcal{D}^\dagger_{s_1}(t)\mathcal{D}_{s_1}\rangle - \sum_{s_1 \in \{\pm\}}\langle \mathcal{D}^\dagger_{s_1}\mathcal{D}_{s_1}(t)\mathcal{D}^\dagger_{-s_1}(t)\mathcal{D}_{-s_1}\rangle\right)$$
$$= M_{XXXX}(t) = \exp(-(\Gamma_S + \Gamma_A)t), \tag{S47a}$$

$$\rho_{SS}(t) = \sum_{s_1, s_2} s_1 s_2 \Big(M_{SSXX}(t)\langle \mathcal{D}_+(t)\mathcal{D}^\dagger_+(t)\mathcal{D}^\dagger_{s_1}\mathcal{D}^\dagger_{s_1}\mathcal{D}_{s_2}\mathcal{D}_{s_2}\rangle$$
$$+ M_{AAXX}(t)\langle \mathcal{D}_-(t)\mathcal{D}^\dagger_-(t)\mathcal{D}^\dagger_{s_1}\mathcal{D}^\dagger_{s_1}\mathcal{D}_{s_2}\mathcal{D}_{s_2}\rangle\Big)$$
$$= \frac{1}{2}\frac{\Gamma_S}{\Gamma_A}\left(1 - e^{-\Gamma_A t}\right)e^{-\Gamma_S t}(1 - \kappa^2) + \frac{1}{2}\frac{\Gamma_A}{\Gamma_S}\left(1 - e^{-\Gamma_S t}\right)e^{-\Gamma_A t}(1 + \kappa^2), \tag{S47b}$$

$$\rho_{AA}(t) = \sum_{s_1, s_2} s_1 s_2 \Big(M_{AAXX}(t)\langle \mathcal{D}_+(t)\mathcal{D}^\dagger_+(t)\mathcal{D}^\dagger_{s_1}\mathcal{D}^\dagger_{s_1}\mathcal{D}_{s_2}\mathcal{D}_{s_2}\rangle$$
$$+ M_{SSXX}(t)\langle \mathcal{D}_-(t)\mathcal{D}^\dagger_-(t)\mathcal{D}^\dagger_{s_1}\mathcal{D}^\dagger_{s_1}\mathcal{D}_{s_2}\mathcal{D}_{s_2}\rangle\Big)$$
$$= \frac{1}{2}\frac{\Gamma_S}{\Gamma_A}\left(1 - e^{-\Gamma_A t}\right)e^{-\Gamma_S t}(1 + \kappa^2) + \frac{1}{2}\frac{\Gamma_A}{\Gamma_S}\left(1 - e^{-\Gamma_S t}\right)e^{-\Gamma_A t}(1 - \kappa^2), \tag{S47c}$$

$$\bar{\rho}_{GG}(t) = M_{GGXX}(t)\left(\sum_{s_1}\langle \mathcal{D}^\dagger_{s_1}\mathcal{D}_{s_1}(t)\mathcal{D}^\dagger_{s_1}(t)\mathcal{D}_{s_1}\rangle - \sum_{s_1}\langle \mathcal{D}^\dagger_{s_1}\mathcal{D}_{s_1}(t)\mathcal{D}^\dagger_{-s_1}(t)\mathcal{D}_{-s_1}\rangle\right)$$
$$= 1 + \frac{\Gamma_A^2 - \Gamma_A\Gamma_S + \Gamma_S^2}{\Gamma_A\Gamma_S}e^{-(\Gamma_S + \Gamma_A)t} - \frac{\Gamma_A}{\Gamma_S}e^{-\Gamma_A t} - \frac{\Gamma_S}{\Gamma_A}e^{-\Gamma_S t}. \tag{S47d}$$

For the solutions of the environment correlation functions appearing in these derivations, we refer to Sec. S3.2.

When initializing within the single-excitation manifold, the equilibration of the phonon bath leads to transitions between the symmetric and antisymmetric state of the system, while the ground state gets occupied due to radiative decay processes. The doubly excited state stays unoccupied. For $\rho_S(0) = |\Psi_S\rangle\langle\Psi_S|$ we find

$$\rho_{SS}(t) = M_{SSSS}(t)\langle \mathcal{D}^\dagger_+\mathcal{D}_+(t)\mathcal{D}^\dagger_+(t)\mathcal{D}_+\rangle + M_{AAAA}(t)\langle \mathcal{D}^\dagger_-\mathcal{D}_-(t)\mathcal{D}^\dagger_-(t)\mathcal{D}_-\rangle$$
$$+ M_{SASA}(t)\langle \mathcal{D}^\dagger_-\mathcal{D}_-(t)\mathcal{D}^\dagger_+(t)\mathcal{D}_+\rangle + M_{ASAS}(t)\langle \mathcal{D}^\dagger_+\mathcal{D}_+(t)\mathcal{D}^\dagger_-(t)\mathcal{D}_-\rangle,$$
$$\rho_{AA}(t) = M_{SSSS}(t)\langle \mathcal{D}^\dagger_+\mathcal{D}_-(t)\mathcal{D}^\dagger_-(t)\mathcal{D}_+\rangle + M_{AAAA}(t)\langle \mathcal{D}^\dagger_-\mathcal{D}_+(t)\mathcal{D}^\dagger_+(t)\mathcal{D}_-\rangle$$
$$+ M_{SASA}(t)\langle \mathcal{D}^\dagger_-\mathcal{D}_+(t)\mathcal{D}^\dagger_+(t)\mathcal{D}_+\rangle + M_{ASAS}(t)\langle \mathcal{D}^\dagger_+\mathcal{D}_-(t)\mathcal{D}^\dagger_+(t)\mathcal{D}_-\rangle,$$
$$\rho_{GG}(t) = M_{GGSS}(t)\langle \mathcal{D}^\dagger_+\mathcal{D}_+\rangle + M_{GGAA}(t)\langle \mathcal{D}^\dagger_-\mathcal{D}_-\rangle. \tag{S48}$$

11

Similarly, we obtain for $\rho_S(0) = |\Psi_S\rangle\langle\Psi_S|$

$$\begin{aligned}
\rho_{SS}(t) =& M_{SSSS}(t)\langle\mathcal{D}^\dagger_-\mathcal{D}_+(t)\mathcal{D}^\dagger_+(t)\mathcal{D}_-\rangle + M_{AAAA}(t)\langle\mathcal{D}^\dagger_+\mathcal{D}_-(t)\mathcal{D}^\dagger_-(t)\mathcal{D}_+\rangle \\
&+ M_{SASA}(t)\langle\mathcal{D}^\dagger_+\mathcal{D}_-(t)\mathcal{D}^\dagger_+(t)\mathcal{D}_-\rangle + M_{ASAS}(t)\langle\mathcal{D}^\dagger_-\mathcal{D}_+(t)\mathcal{D}^\dagger_-(t)\mathcal{D}_+\rangle, \\
\rho_{AA}(t) =& M_{SSSS}(t)\langle\mathcal{D}^\dagger_-\mathcal{D}_-(t)\mathcal{D}^\dagger_-(t)\mathcal{D}_-\rangle + M_{AAAA}(t)\langle\mathcal{D}^\dagger_+\mathcal{D}_+(t)\mathcal{D}^\dagger_+(t)\mathcal{D}_+\rangle \\
&+ M_{SASA}(t)\langle\mathcal{D}^\dagger_+\mathcal{D}_+(t)\mathcal{D}^\dagger_-(t)\mathcal{D}_-\rangle + M_{ASAS}(t)\langle\mathcal{D}^\dagger_-\mathcal{D}_-(t)\mathcal{D}^\dagger_+(t)\mathcal{D}_+\rangle, \\
\rho_{GG}(t) =& M_{GGSS}(t)\langle\mathcal{D}^\dagger_-\mathcal{D}_-\rangle + M_{GGAA}(t)\langle\mathcal{D}^\dagger_+\mathcal{D}_+\rangle.
\end{aligned} \tag{S49}$$

The environment correlations are derived in Sec. S3.2 and listed in Eqs. (S40) & (S41), while for the ground state we make use of Eq. (S25a) together with $\langle D_j\rangle = \kappa$, $j = 1, 2$.

Inserting these results into the above equations we find the occupations the system evolving according to

$$\begin{aligned}
\rho_{SS}(t) =& \rho_{SS}(0)\left(e^{-\Gamma_S t}\mathcal{E}_{++}(t) + e^{-\Gamma_A t}\mathcal{E}_{-+}(t) + e^{-\gamma t}\mathcal{E}_0(t)\right) \\
&+ \rho_{AA}(0)\left(-e^{-\gamma t}\mathcal{E}_0(t) + e^{-\Gamma_S t}\mathcal{E}_{--}(t) + e^{-\Gamma_A t}\mathcal{E}_{+-}(t)\right),
\end{aligned} \tag{S50a}$$

$$\begin{aligned}
\bar{\rho}_{AA}(t) =& \rho_{SS}(0)\left(e^{-\Gamma_S t}\mathcal{E}_{+-}(t) + e^{-\Gamma_A t}\mathcal{E}_{--}(t) - e^{-\gamma t}\mathcal{E}_0(t)\right) \\
&+ \rho_{AA}(0)\left(e^{-\gamma t}\mathcal{E}_0(t) + e^{-\Gamma_S t}\mathcal{E}_{-+}(t) + e^{-\Gamma_A t}\mathcal{E}_{++}(t)\right)
\end{aligned} \tag{S50b}$$

with

$$\mathcal{E}_0 = \frac{\kappa^4}{2}\sinh(2\Re[\Phi(t)]), \tag{S50c}$$

$$\mathcal{E}_{s_1 s_2} = \frac{1}{4}\left[1 + s_1\kappa^2 + s_2\left[\kappa^4\cosh(2\Re[\Phi(t)]) + s_1\kappa^2\cos(2\Im[\Phi(t)])\right]\right]. \tag{S50d}$$

## S4 Numerically exact simulations

To compare our analytical calculations with exact numerics, and to obtain numerical results for pulsed laser driving, we employ a numerically exact process tensor method [9, 12, 15]. This method allows us to calculate the influence of an arbitrary boson environment on a system independently from its internal dynamics, store this influence in the process tensor and afterwards combine multiple process tensors to account for possible non-additive effects between environments.

To account for the phonon environment contributions we assume individual environments for both emitters (cf. Ref. [14]). As described in the main text, we take the example of GaAs quantum dots with a diameter of about 4 nm, for which the spectral density can be calculated from first principles [2, 3]:

$$J_{PN}(\omega) = \frac{\omega^3}{2\mu\hbar c_s^5}\left(D_e e^{-\frac{\omega^2}{\omega_e^2}} - D_h e^{-\frac{\omega^2}{\omega_h^2}}\right)^2, \tag{S51}$$

with the mass density $\mu = 5370$ kg/m$^3$, the electron (hole) deformation potential $D_e = 7.0$ eV ($D_h = -3.5$ eV), the electron (hole) cutoff frequency $\omega_e = 2.9$ meV ($\omega_h = 4.4$ meV) and the speed of sound $c_s = 5110$ m/s.

To confirm our analytical calculations for the single two-level system, we include the electromagnetic environment numerically exactly using a process tensor, as well. For this we describe the coupling to the electromagnetic field via the Rabi Hamiltonian

$$H = \hbar\omega_X \sigma^+\sigma^- + \hbar \sum_{\mathbf{k}} \omega_{\mathbf{k}} a_{\mathbf{k}}^\dagger a_{\mathbf{k}} + \hbar\sigma^x \sum_{\mathbf{k}} g_{\mathbf{k}}(a_{\mathbf{k}}^\dagger + a_{\mathbf{k}}). \tag{S52}$$

A Hamiltonian of this form allows for the calculation of the process tensor using efficient divide-and conquer methods [15]. This, however, comes to the price of needing to resolve the resulting high-frequency oscillations of the counter-rotating terms. The spectral density for the electromagnetic environment reads

$$J_{PT}(\omega) = \frac{\gamma}{2\pi}\frac{1}{1+e^{-\omega/w}}\frac{1}{1+e^{(\omega_{max}-\omega)/w}}, \tag{S53}$$

and is approximately box-shaped with a broad flat region around $\omega_{max}/2$ and rounded cutoffs at $\omega = 0$ and $\omega = 2\omega_{max}$. For appropriate convergence parameters, this spectral density describes radiative decay with rate $\gamma$, if $\omega_X$ is far away from the cutoffs.

For sufficiently large $\omega_X$, the counter-rotating contributions of the Rabi Hamiltonian (S52) are suppressed and the dynamics are equivalent to the one presented in the main section. However, the finite transition frequency $\omega_X$ introduces fast oscillations of the coherences, which we remove by post-processing our numerical results. Notably, convergence for the coherences is achieved by moving the upper frequency cutoff to values around $\omega_{max} = 20\omega_X$. Furthermore, we choose $\omega_X = 1000\gamma$, $w = 100\gamma$, and $\gamma = 0.005 \frac{1}{\text{ps}}$.

To determine the process tensor for superradiant decay, we use the same spectral density and the Hamiltonian

$$H_S + H_{PT} = \hbar\omega_X \sum_{j=1,2} \sigma_j^+\sigma_j^- + \hbar \sum_{\mathbf{k}} \omega_{\mathbf{k}} a_{\mathbf{k}}^\dagger a_{\mathbf{k}} + \hbar(\sigma_1^x + \sigma_2^x) \sum_{\mathbf{k}} g_{\mathbf{k}}(a_{\mathbf{k}}^\dagger + a_{\mathbf{k}}). \tag{S54}$$

We have confirmed that virtually identical results are obtained when the effects of the photon environment are described using a Lindblad superoperator

$$\mathcal{L}[\sigma_S^-]\rho = 2\gamma\left(\sigma_S^- \rho \sigma_S^+ - \frac{1}{2}[\sigma_S^+\sigma_S^-, \rho]_+\right).\mathcal{L}[\sigma_S^-]\rho = 2\gamma\left(\sigma_S^- \rho \sigma_S^+ - \frac{1}{2}[\sigma_S^+\sigma_S^-, \rho]_+\right). \tag{S55}$$


\end{document}